\documentclass[10pt]{iopart}

\usepackage{graphicx}
\expandafter\let\csname equation*\endcsname\relax
\expandafter\let\csname endequation*\endcsname\relax
\usepackage{mathptmx} 
\usepackage{newtxtext, newtxmath} 
\usepackage{verbatim}
\usepackage{mathrsfs}
\usepackage{color}
\usepackage{bm} 
\usepackage{xspace}
\usepackage{hyperref}
\usepackage{multirow}
\usepackage{color}
\usepackage{float}
\usepackage{bbold}
\usepackage{tabu}
\usepackage{textcomp}
\usepackage{dcolumn}
\usepackage{diagbox} 

\begin{document}

\title[]{Electronic structure of lattice relaxed alternating twist t$N$G-multilayer graphene: from few layers to bulk AT-graphite}

\author{Nicolas Leconte$^1$, Youngju Park$^1$, Jiaqi An$^1$, Appalakondaiah Samudrala$^1$, Jeil Jung$^{1,2}$}
\address{$^1$ Department of Physics, University of Seoul, Seoul 02504, Korea}
\address{$^2$ Department of Smart Cities, University of Seoul, Seoul 02504, Korea}
\ead{jeiljung@uos.ac.kr}

\begin{abstract}
We calculate the electronic structure of AA$^\prime$AA$^\prime$\dots\-stacked
alternating twist $N$-layer (t$N$G) graphene for $N = $3, 4, 5, 6, 8, 10, 20 layers and bulk alternating twist (AT) graphite systems where the lattice relaxations are modeled by means of molecular dynamics simulations.
We show that the symmetric AA$^\prime$AA$^\prime$\dots stacking is energetically preferred among all interlayer sliding geometries for progressively added layers up to $N=6$. 
Lattice relaxations enhance electron-hole asymmetry, and reduce the magic angles with respect to calculations with fixed tunneling strengths that we quantify from few layers to bulk AT-graphite.
Without a perpendicular electric field, the largest magic angle flat-band states locate around the middle layers following the largest eigenvalue eigenstate in a 1D-chain model of layers, while the density redistributes to outer layers for smaller magic twist angles corresponding to higher order effective bilayers in the 1D chain. 
A perpendicular electric field decouples the electronic structure into $N$ Dirac bands with renormalized Fermi velocities with distinct even-odd band splitting behaviors, showing a gap for N=4 while for odd layers a Dirac cone remains between the flat band gaps. The magic angle error tolerance estimated from density of states maxima expand progressively from $0.05^{\circ}$ in t2G to up to $0.2^{\circ}$ in AT-graphite, hence allowing a greater flexibility in multilayers. 
Decoupling of t$N$G into t2G using effective interlayer tunneling proportional to the eigenvalues of a 1D layers chain allows to map t$N$G-multilayers bands onto those of periodic bulk AT-graphite's at different $k_z$ values. We also obtain the Landau level density of states in the quantum Hall regime for magnetic fields of up to 50~T and confirm the presence of nearly flat bands around which suppressed density of states gap regions can develop by applying an electric field in $N > 3$ systems.
\end{abstract}
\maketitle
\ioptwocol

\section{Introduction}


Twisted graphene systems have emerged as the prime target and candidate for finding 
correlated electronic phenomena in 2D materials including flat band superconductivity~\cite{cao2018,Park2021,Hao2021,li2019, Cea2019, PhysRevResearch.2.022010}.
Research has expanded beyond the initial investigation 
of twisted bilayer graphene (t2G)~\cite{cao2018}
to include a variety of other layered moire materials that 
have shown hints of similar physics~\cite{1901.09356,  Liu2020, 2203.09188}.
Alternating twist $N$-layer graphene (t$N$G) systems are particularly interesting 
for exploring correlation physics in light of recent experiments that have revealed
flat band superconductivity for $N = 3,\, 4,\, 5$ systems~\cite{jeongminpark2021, Hao2021, jeongminpark2022}. 
%

In this work we studied the lattice relaxation effects in t$N$G systems 
by means of molecular dynamics calculations. 
The topmost layer was allowed to slide through the different stacking configurations
in order to show that t$N$G with lowest total energies correspond to systems where local AA stacking are vertically aligned.
We then mainly targeted our study on the largest magic angles for each t$N$G multilayer
with most stable AA$^\prime$AA$^\prime$\dots\-stacking and showed
how the associated flat bands vary with experimental conditions like twist angle or tunneling strengths, electric fields and magnetic fields. 
In an effort to systematically improve prediction reliability and accuracy the relaxed atomic structures are obtained through real-space molecular dynamics calculations and the electronic structure analysis uses models that are informed by density functional theory~\cite{PhysRevB.96.195431}.

The atomic relaxation is shown to affect the band structure by enhancing the asymmetry between the conduction and valence bands, separating the low energy bands from the higher bands at the $\Gamma$ point as found in t2G, and allows 
to determine more clearly the flat band magic angles where the density of states (DOS) become maximum. From DOS calculations we observe a small decrease in the actual magic angle values when the relaxation effects are considered. 
We show that the charge associated to the flat bands are preferentially located around the middle layers of t$N$G for the largest magic angles, following the eigenstates of a 1D chain model of layers whose associated eigenvalues are proportional to the interlayer tunneling of effective bilayers.
The decomposition of t$N$G into effective bilayers allows a mapping between t$N$G and bulk alternating twist (AT) graphite at selected $k_z$ values.
%
In the presence of perpendicular electric fields we see energy split Dirac cones near $K$ and $K^{\prime}$ points in the moire Brillouin zone in even-$N$ multilayer systems while one unshifted Dirac cone remains in odd-$N$ systems. This distinct even-odd behavior allows the development of band gaps in a t4G system near its largest magic angles and a Dirac cone is expected at charge neutrality for odd number of layers. 
%
Based on Landau level density of states calculations for up to $N = 6$ layers and magnetic fields of up to 50~Teslas
in the quantum Hall regime we show the existence of constant energy DOS peaks associated to the nearly flat bands, 
and suppressed DOS gapped regions for $N>3$ systems when we apply electric fields that can decouple the different effective Dirac cones within the t$N$G.

The paper is organized as follows. In Sect.~\ref{methodsSection}, 
we introduce the continuum and tight-binding Hamiltonian models used for our calculations. 
In Sect.~\ref{resultSection}, we discuss the energetic stability of the system, the reduction of the magic angle predictions due to relaxation effects, as well as other observations made based on the DOS and electronic band structure for zero and finite magnetic fields, and then we elaborate on the finite system to bulk mapping. 
The summary and conclusions are presented in Sect.~\ref{conclusionsSection}.

\section{Methods}
\label{methodsSection}
Here we outline our calculation methods based on the continuum moire bands theory~\cite{Bistritzer:2011ho} with fixed interlayer tunneling parametrizations of Ref.~\cite{chebrolu2019}, and real-space tight-binding calculations~\cite{TRAMBLYDELAISSARDIERE:2013fa} using the atomic and electronic structure calculation methods proposed in Ref.~\cite{1910.12805}.

\subsection{Continuum model calculations}
The continuum model Hamiltonian for t$N$G that we use 
is based on the moire bands theory for t2G~\cite{Bistritzer:2011ho}
and extensions thereof~\cite{jung2014}
using the parametrization in Ref.~\cite{chebrolu2019}
for unequal interlayer inter-sublattice and intra-sublattice tunneling constants $\omega$ and $\omega^{\prime}$.
The Hamiltonian reads
\begin{equation}
\begin{aligned}
    {\cal H}_{\bm{k}} = 
    \begin{pmatrix} H_{\bm{k}}^-   & { T}(\bm{r}) & 0   & \cdots \\ 
    { T}^{\dagger}(\bm{r}) & H_{\bm{k}'}^+  &  {T}^{\dagger}(\bm{r})  &  \cdots \\
    0 & { T}(\bm{r}) & H_{\bm{k}}^- &  \\
    \vdots & \vdots  & \vdots   & \ddots \end{pmatrix} +  \Delta \epsilon ,
\end{aligned}
\label{Eq:continuumHamil}
\end{equation}
where the diagonal blocks contain each alternating $\pm$ twist graphene layer's Dirac cones, 
\begin{equation}
\begin{aligned}
H_{\bm{k}}^{\pm} &=
    \hbar \upsilon_F 
    \begin{pmatrix} 0 & \pi_{\bm{k}}^{\dagger} e^{\mp i \theta/2} \\
    \pi_{\bm{k}}e^{\pm i \theta/2} & 0  \end{pmatrix}.
\end{aligned}
\end{equation}
Here $\pi_{\bm{k}} = {\bar{\bm{k}}}_x + i {\bar{\bm{k}}}_y$ where the momentum vector ${\bar{\bm{k}}}$ is measured from the rotated Dirac cones. 
We use the Fermi velocity at a single Dirac cone to be
$\upsilon_F = (\sqrt{3}a/2 \hbar)|t_{0}|\approx 10^{6}~\rm m/s$ with the effective nearest hopping parameter $t_0=-3.1~\rm eV$ and a graphene lattice constant $a=2.46$~\AA.
The off-diagonal blocks are the interlayer tunneling between the twisted layers,
\begin{equation}
\begin{aligned}
{ T}(\bm{r}) &= \sum_{j=0,\pm} e^{-i \bm{Q}_j \cdot \bm{r}}
~\left[\begin{pmatrix}
\omega^{\prime} &\omega~ e^{-i\phi_j}   \\  
\omega~ e^{i\phi_j}  & \omega^{\prime}
\end{pmatrix}~e^{-i \bm{{\cal G}}_j \cdot \bm{\tau}_s}\right]
\end{aligned}
\label{Eq:tunneling}
\end{equation}
where $\phi_j = (2\pi/3)j$, $\bm{Q}_{j} = {\rm Rot}_{\phi_j}([K\theta(0,-1)])$ connecting the two adjacent layers' Dirac points, where we use $K=4\pi/3a$.
The tunneling amplitudes of $\omega = \omega_{BA^{\prime}} = t_1 /3$ and $\omega^{\prime} =\omega_{AA^{\prime}}= (-0.1835 {t_1}^2 + 1.036 {t_1} - 0.06736)/3$ are estimated from the EXX+RPA fitting values~\cite{chebrolu2019} with $t_1=0.36~\rm eV$, 
and the different $\omega$ and $\omega^{\prime}$ values reflect the effect of out-of-plane atomic relaxation.
The $\bm{{\cal G}}_{j}$ that correspond to $\bm{Q}_{j}$ are the reciprocal lattice vectors of the original graphene, and $\bm{\tau}_s$ denotes the relative sliding displacement before the rotation that we have applied for the topmost pair of layers.
The bulk alternating twist AT-graphite Hamiltonian at $k_z$ reduces to the following form
where the interlayer tunneling matrix acquires a z-dependent phase term as follows
\begin{equation}
\begin{aligned}
    {\cal H}_{\bm{k}}({k}_z) = 
    \begin{pmatrix} H_{\bm{k}}^-   & 2 { T}(\bm{r})  \cos( {k}_z c_0) \\ 
   2 { T}^{\dagger}(\bm{r}) \cos( {k}_z c_0) & H_{\bm{k}'}^+  \end{pmatrix} ,
\end{aligned}
\label{Eq:continuumHamilbulk}
\end{equation}
with $c_0=3.35$~\AA\ 
a constant interlayer separation distance, and we assume that the vertical size of a unit cell is $2c_0$.
The last term in Eq.~(\ref{Eq:continuumHamil}) captures the effect of an electric field as
\begin{equation}
    \Delta \epsilon = {\rm diag}\left(-\frac{N-1}{2},-\frac{N-1}{2}+1,\dots,+\frac{N-1}{2}\right) \cdot \frac{\Delta V}{N-1}\mathbb{1}.
    \label{electricEq}
\end{equation}
Both intra- and interlayer relaxation affect the momentum-space tunneling amplitudes $\omega^{(\prime)}$ in Eq.~(\ref{Eq:tunneling}).
We estimate the deformed tunneling amplitudes to $\bm{{\cal G}}_j$ at 
\begin{equation}
\begin{aligned}
\omega_{XX^{\prime}}(\bm{{\cal G}}_j) 
&= H_{XX^{\prime}} \left( \bm{K}:\bm{{\cal G}}_j\right) \\
&= \frac{1}{A_M}\int_{A_M}d\bm{r}~\left[
 e^{i\bm{{\cal G}}_j\,\cdot\,\bm{d}(\bm{r})} ~ H_{XX^{\prime}} \left( \bm{K}:\bm{d}(\bm{r}) \right)\right]
\end{aligned}
\label{Eq:omega}
\end{equation}
starting from the Wannier representation Bloch-band Hamiltonian 
$H_{XX^{\prime}} \left( \bm{K}:\bm{d}(\bm{r}_i) \right)
= \sum_{j}\ t_{ij}^{\rm inter}\ e^{i\bm{K}\cdot r_{ij}}$~\cite{1910.12805,jung2014}
where $X^{(\prime)}$ is A or B sublattice of bottom(top) layer and the 
interlayer hopping parameter $t_{ij}^{\rm inter}$ is defined in Eq.~(\ref{STCtunneling}).
The integration runs over a moire commensurate cell of area $A_M$, and
we include both in-plane($\bm{u}$) and out-of-plane($h$) stacking deformation $\bm{d}(\bm{r}) =\bm{d}_0(\bm{r}) +\bm{u}(\bm{r}) + h(\bm{r})\hat{z}$ where 
$\bm{d}_0(\bm{r})$ is the rigid relative stacking information~\cite{10.1103/physrevb.96.085442}.

\subsection{Tight-binding calculations}

The tight-binding (TB) Hamiltonian in the basis $| i \rangle $ of localized states at site 
$i$ is given by~\cite{McKinnon1993} 
\begin{equation}
    \hat{H} = \sum_{i}^{n_{at}} \epsilon_i  |i \rangle \langle i | + \sum_{i,j}^{n_{at}} t_{ij} |i \rangle  \langle j |
\end{equation}
with eigenfunctions
\begin{equation}
    | k \rangle = \frac{1}{\sqrt{n_{at}}} \sum_j^{n_{at}} e^{i {\bm k} \cdot {\bm r}_j} | j \rangle
\end{equation}
with $n_{at}$ the number of atoms and
where ${\bm k} = (k_x,k_y)$ for the 2D systems and  ${\bm k} = (k_x,k_y,k_z)$ for the bulk calculations. The $\epsilon_i$ are the onsite energies provided by the F2G2 model of graphene~\cite{jung2013} that can 
additionally take potential energy shifts due to an electric field as in Eq.~(\ref{electricEq}).
The hopping terms $t_{ij} = t_{ij}^{\rm intra} + t_{ij}^{\rm inter}$ consist of intralayer $t_{ij}^{\rm intra}$ terms given by the F2G2 model and the interlayer $t_{ij}^{\rm inter}$ distance-dependent model of Ref.~\cite{1910.12805} 
with a scaling parameter $S$ that reads
\begin{equation}
   t^{\rm inter}_{ij} = S \, \exp \left[ \frac{c_{ij} -p }{q} \right]  t^{\rm inter}_{{\rm TC}, ij}
\label{STCtunneling}
\end{equation}
where $S$ controls the magic angle value and $p=3.25$~\AA\ and $q=1.34$~\AA\ control the interlayer distance-dependent fitting of the tunneling at the K-point and
\begin{equation}
    t_{ {\rm TC}, \, ij} = V_{pp\pi}(r_{ij}) \left[1 - \left(\frac{ c_{ij} }{r_{ij}} \right)^2 \right] + V_{pp\sigma}(r_{ij}) \left(\frac{ c_{ij} }{r_{ij}} \right)^2
\label{KoshinoSREquation}
\end{equation}
where
\begin{equation}
    V_{pp\pi}(r_{ij}) = V_{pp\pi}^0 \exp\left(-\frac{r_{ij}-a_0}{r_0}\right)
    \label{vpppiEq}
\end{equation}
and
\begin{equation}
    V_{pp\sigma}(r_{ij}) = V_{pp\sigma}^0 \exp\left(-\frac{r_{ij}-c_0}{r_0}\right)
\label{vppsigmaEq}
\end{equation}
with the interlayer distance $c_0 = 3.35$~\AA, 
the rigid interatomic carbon distance in graphene $a_0 = 1.42$~\AA, the transfer integral between nearest-neighbor atoms $V_{pp\pi}^0 = -2.7$~eV,  the transfer integral between two vertically aligned atoms $V_{pp\sigma}^0 = 0.48$~eV,
the decay length of the transfer integral set to $r_0 = 0.184 a$ such that the next-nearest intralayer coupling becomes 0.1 $V_{pp\sigma}^0$ and the magnitude of the interatomic distance $r_{ij} = \left| {\bm r}_{ij} \right|$. The cutoff for this distance-dependent model is finally set to  $4.9$~\AA\ beyond which additional contributions do not affect the observables anymore~\cite{laissardire2012}. The coefficient $S = 0.895$ in Eq.~(\ref{STCtunneling}) is obtained using
\begin{equation}
S = \frac{\theta_1 |t_\text{eff}|}{\omega} s = C_1 s
\label{Sfact}
\end{equation}
where $s$ contains the relaxation-specific effects and are calibrated to give the magic angle 
at $\theta_1 = 1.08^\circ$ using $t_\text{eff} = -3.1$~eV and $\omega = 0.11$~eV based on 
rigid bilayer graphene calculations~\cite{Chittari2018}. 
%
In the so called continuum chiral model for multilayer t$N$G systems~\cite{khalaf2019}, 
where the same sublattice tunneling between consecutive layers is set to zero, 
the analytical flat band angles are given by 
\begin{equation}
\theta^{(N)} = \omega / (\upsilon_{\rm F} k_D \alpha^{(N)}_k)
\label{hierarchy}
\end{equation}
through the dimensionless constant $\alpha^{(N)}_k = \alpha^{(2)} / \lambda_{k}$ 
for non-negative values of $\lambda_k$ when $k = 1,\dots,N_e$ and $N_e = [N / 2]$ is 
the number of t2G bilayers forming the alternating twist multilayer system such that $N = 2 N_e$ for even-layered 
systems and $N = 2 N_e + 1$ for odd-layered systems. 
The eigenvalues and eigenvectors of an $N$ layers 1D-chain with unity hopping energy are given by
\begin{eqnarray}
     \lambda_k &=& 2 \cos{\left( \kappa_{ k } \right)},  \label{1dchainEq}  \\
     \psi_{k}( \ell ) &=& \sqrt{\frac{2}{N+1}}  
\sin \left( \kappa_k\ \ell \right),  \label{eigenstates}
\end{eqnarray}
where $\kappa_{k} = \pi k / (N + 1)$ and $\ell = 1, 2, \hdots N$ is the layer index.

For the electronic band structure calculations we construct commensurate cells that contain the smallest number of atoms whose moire period equals the period of the commensuration cell. 
Since the magic angle of a given system and the interlayer tunneling
scaling parameter $S'$ satisfy the relation $S'/\theta' = S/\theta$
the band flattening can be achieved at different physical magic twist angles if we modify the interlayer coupling strength. 
We use the scalability of the electronic structure with interlayer tunneling 
to calculate the band structure of a given effective twist angle $\theta_{\rm eff}$ 
based on a commensurate reference twist angle $\theta_{\rm ref}$. 
The updates in $S^{\prime}$ for changes in $\delta \theta$, $\delta \upsilon_{\rm F}$, 
or $\delta \omega$ parameters can be introduced through
\begin{equation}
S^{\prime} = C^{\prime}_1 s = C_1 \left( 1 + \frac{\delta \theta}{\theta_{\rm ref}} \right) \left( 1 + \frac{\delta \upsilon_{\rm F}}{\upsilon_{\rm F}} \right) \left( \frac{\omega}{\omega + \delta \omega} \right) s.
\label{effectiveS}
\end{equation}
and the electronic band structures at effective twist angles can be obtained through
\begin{equation}
\theta_{\rm eff} = \theta_{\rm ref} - \delta \theta
= \frac{S}{S^{\prime}} \theta_{\rm ref} 
\label{effTheta}
\end{equation}
where $\theta_{\rm ref}$ is a commensurate superlattice angle close to the value of the effective $\theta_{\rm eff}$ that we want to calculate. 
The specific choice of $S^\prime$ for each system is summarized in Table~\ref{tab:systemparameters} from the Appendix that minimizes the bandwidth 
\begin{equation}
W_{\Gamma} = E_{\rm cond}(\Gamma) - E_{\rm val}(\Gamma)
\end{equation}
that we defined subtracting the maximum of the lowest conduction band and the minimum of the highest valence band at the $\Gamma$-point. We verified that our estimate of the magic angle remains the same if we choose the valence band reference value to be at the M-point. For most of the quantitative magic angle analysis presented in Sect.~\ref{DOSSect} we have based our band flatness criterion instead on the DOS maxima peaks as a function of $\theta$ thus leading to small differences for $\theta_{\rm eff}$ obtained from the electronic band structures and the DOS-inferred $\theta$-values among $E$ values near charge neutrality
\begin{equation}
     \theta_{\rm M} = {\rm argmax}_{\theta} \,\, {\rm DOS} (E,\theta).
\end{equation}
The DOS calculations have been carried out through direct diagonalization when allowed by the system sizes and by using the Lanczos method when we need to calculate very large systems with millions of atoms
\begin{eqnarray}
   {\rm DOS}_{\eta}(E) 
   &=&  - \frac{1}{\pi}\Im m  \left\langle \varphi_{RP} \middle| \frac{1}{E+i\eta - \tilde{\hat{H}}} \middle|\varphi_{RP} \right\rangle  
    \label{eqDos}
\end{eqnarray}
where RP refers to a random phase being used to approximate the trace of large matrices~\cite{lanczos} and $\eta$ is the broadening factor that allows to control the energy resolution.
We finally note that for the magnetic field calculations, we use a Peierls phase substitution in the hopping terms as discussed in Ref.~\cite{PhysRevB.103.045402}.

\subsection{Molecular dynamics simulations}

The value $S=0.895$ used in Eq.~(\ref{STCtunneling}) as well as the modified $S^\prime$ values from Eq.~(\ref{effectiveS}) are suitable when using the REBO2 force-field~\cite{Brenner_2002} for the intralayer interactions and the EXX-RPA-informed~\cite{PhysRevB.96.195431} DRIP~\cite{PhysRevB.98.235404} interlayer potential when performing the MD calculations. Specifically, 
we performed the MD calculations using LAMMPS~\cite{Plimpton1995} with the FIRE minimization scheme~\cite{PhysRevLett.97.170201} with a time step of $0.001$ ps and a stopping tolerance on the forces of $0.001$ eV/\AA. Interlayer interactions are only included between adjacent layers. 
We note that the band separation between the flat bands and the rest of the spectrum is over-estimated when using the present REBO2 potential~\cite{1910.12805}, while better mechanical predictions can be obtained using the computationally more demanding GAP$_{20}$ potential~\cite{Rowe2020}.

\section{Atomic and electronic structure of t$N$G} 
\label{resultSection}

The alternating twist t$N$G multilayers with equal twist angle magnitudes
give rise to moire patterns at each interface that have the same period and whose angles are aligned. 
The resulting moire patterns can still have different interlayer stacking 
but here we focus our atomic and electronic structure study 
mainly on the properties of AA$^\prime$AA$^\prime$\dots\ structures 
where all graphene layers are exactly on top of each other before
rotating the layers, based on the energetic considerations from Sect.~\ref{energeticsSect}.

\subsection{Sliding energetics}
\label{energeticsSect}

\begin{figure*}[tbhp]
\begin{center}
\includegraphics[width=1.0\textwidth]{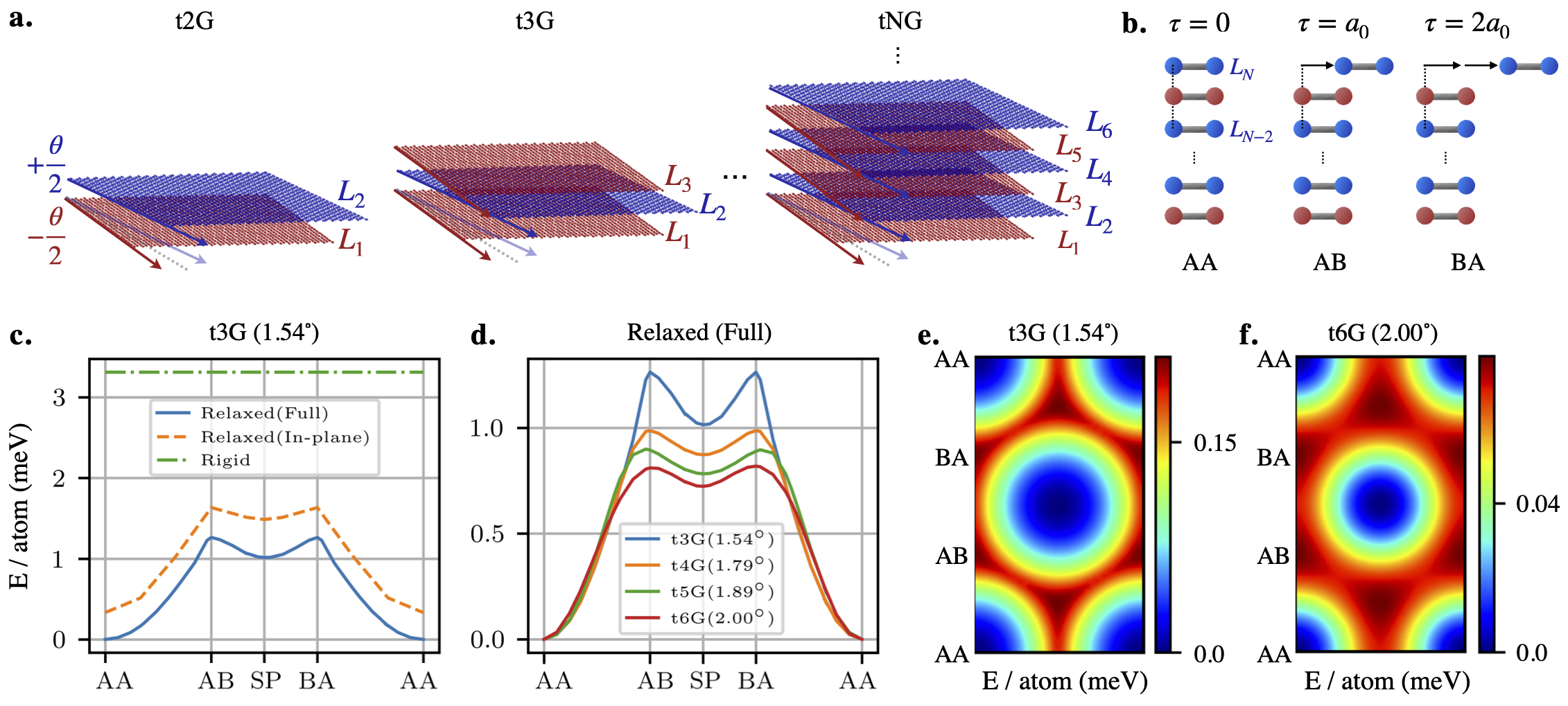}
\caption{(color online) 
{\bf a.} Schematic of alternating twist t2G, t3G, t6G multilayers and beyond numbered from bottom to top. 
The $\mp \theta/2$ alternating twists are shown around a given reference axis represented in gray. 
{\bf b.} Sliding vectors ${\bm \tau}$ prior to twisting the layers define the local stacking.  We always slide the uppermost layer $N$ with respect to layer $N-2$ through AA, AB and BA stackings defined in the same way as non-twisted bilayer graphene systems.
{\bf c.} Energy per atom for the t3L system where we gradually include additional degrees of freedom from the rigid (dashed-dotted green) to the in-plane relaxed system (dashed orange) and the fully relaxed system (solid blue). We set the energy of the AA stacking of the fully relaxed system to zero. {\bf d.} Comparison of the fully relaxed structures with respect to sliding when increasing the number of layers. The angles corresponding to each curve are the ones used for the smallest commensurate cell close to the one of the magic angle of each system, \textit{i.e.} $\theta_\text{cell} = 1.54$, $1.79$, $1.89$ and $2.00^\circ$, respectively. For all systems, the most stable $AA$ structure corresponds to the situation where layer $N$ has zero sliding ${\bm \tau}$ with respect to layer $N-2$. The panels {\bf e.} and {\bf f.} correspond to the same data as panel {\bf d.} but for the full sliding range, for t3G and t6G respectively.}
\label{energetics}
\end{center}
\end{figure*}

Electronic structure studies presented so far have often modeled the simplest AA$^\prime$AA$^\prime$\dots\ structure for t$N$G in Fig.~\ref{energetics}{\bf a.} that does not have any sliding between consecutive layers.
Calculations of sliding energetics for t3G based on molecular dynamics (MD) calculations have shown indeed this is the case for trilayers~\cite{Carr2020}.
%
Here, we extend the calculations for aligned moire angles but variable sliding geometries 
for t$N$G multilayers of up to $N = 6$ layers and verify that a similar behavior takes place for multilayers beyond $N=3$. In our setup we assume that $N=1$ up to $N-1$ do not include any sliding while layer $N$ is translated by ${\bm \tau}$ with respect 
to layer $N-2$ in the usual AA, AB, and BA configurations in Fig.~\ref{energetics} {\bf b.}. 
The corresponding energetics are illustrated in panels {\bf c.}, {\bf d.}, {\bf e.} and {\bf f.}.
In Fig.~\ref{energetics}{\bf c.} we show the total structural energy per atom as a function of stacking for t3G and set the AA-stacking energy at zero.
Comparison between the total energies of rigid (dashed-dotted green), in-plane only (dashed orange) and fully relaxed (solid blue) systems indicates that the largest energy gain happens thanks to the reduction of potential energy when the atoms are allowed to rearrange along the in-plane direction. 
As expected, the rigid configuration does not show any sliding ${\bm \tau}$-dependence. 
The clear reduction in the total energy at the AA stacked regions for the fully relaxed system of about 1~meV/atom on average indicates that this stacking is favored \textit{vs} the AB or BA or the intermediate saddle point SP stacking.
The shape of the energy barriers of in-plane relaxation-only results are nearly the same as for the fully relaxed system, thus highlighting the relative unimportance of the stacking-induced corrugation of the layers,
implying that the energy minimization of the additional graphene layers will mainly depend on the relative moire pattern sliding between the nearest two layers that form the commensurate patterns.
In panel {\bf d.} we show similar ${\bm \tau}$-dependent total energies for $N=3$, $4$, $5$ and $6$ layers-t$N$G at their respective magic angle twists. 
On the right panels in Fig.~\ref{energetics}{\bf e.} and {\bf f.} we show this same total energy landscape as a 2D colormap for all possible slidings for the t3G and the t6G systems. Here, we see that the energetically unfavorable stackings can relax into the most stable AA$^\prime$AA$^\prime$\dots\-stacked system through an energy barrier-free path.
We can gain insight about the relaxation process if we distinguish the total equilibrium interatomic energy   
by breaking it down into an elastic intralayer contribution that rises to resist deformation, 
and the potential interlayer interaction that reduces the total energy of the system by accommodating to the lowest energy stacking. 
In the MD calculations the elastic and potential energy contributions can be obtained during postprocessing,
see Eqs.~(\ref{elastic}, \ref{potential}) for their definitions in the Appendix.
We note that for the AA stacking the total energy reduction due to the potential energy minimization can become almost an order of magnitude more important than the elastic energy penalty, see Fig.~\ref{energetics3L} in the Appendix for illustration on t3G and t4G and further discussions on characteristic sublattice dependent chiral signatures.
For the total potential energy map, the AA stacking shows the highest binding energy, or conversely the lowest free energy; the energetically favorable and unfavorable regions can thus be fully inferred from the potential energy maps.

\subsection{Magic angles based on DOS calculations}
\label{DOSSect}

\begin{figure}[tbhp]
\begin{center}

\includegraphics[width=0.9\columnwidth]{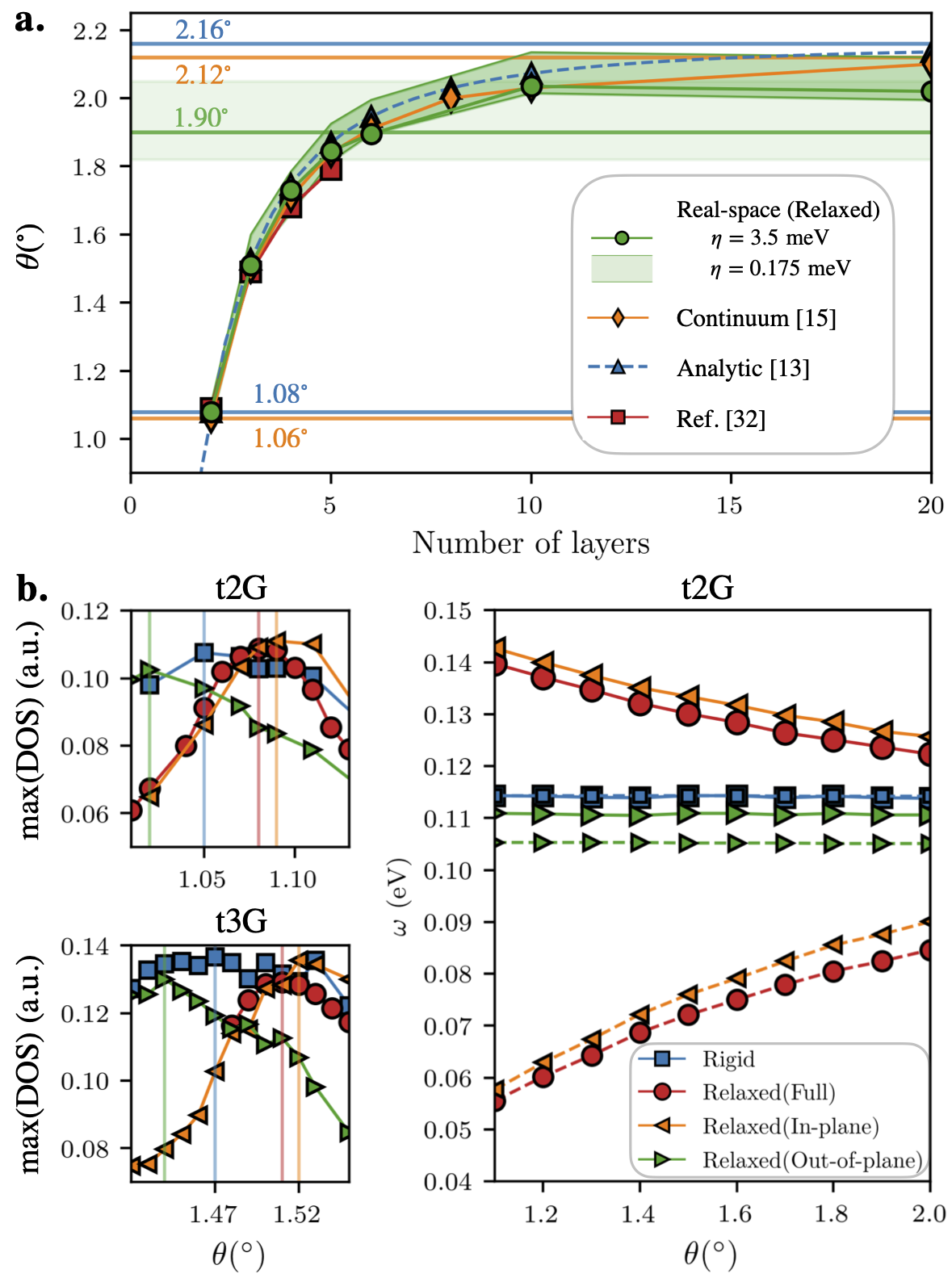}
\caption{(color online) 
{\bf a.} Magic angles as a function of layer number $N$ from calculations based on 
our lattice relaxed (green), lattice rigid (purple), fixed tunneling continuum (orange) from Ref.~\cite{chebrolu2019}, fixed tunneling analytic (blue) from Ref.~\cite{khalaf2019}, and
relaxed continuum model (red) from Ref.~\cite{harvard2021}.
Analytically predicted values of the largest magic angles are obtained from the magic angle hierarchy from Ref.~\cite{khalaf2019} using Eq.~\ref{hierarchy} and our numerically calculated values are obtained using large scale DOS calculations using Eq.~\ref{eqDos}. The green-shaded horizontal line and region correspond to the $\eta=3.5$ and $\eta=0.175$ meV bulk AT-graphite predictions from the real-space calculations. 
We refer to Figs.~\ref{extractMagicAngles_0_001} 
regarding the method used to extract the values reported in this figure.
{\bf b.} t2G (upper-left) and t3G (lower-right) magic angle predictions for the rigid (blue), in-plane-only relaxed (orange), out-of-plane-only relaxed (green) and fully relaxed configurations (red) where the vertical lines indicate the magic angle as predicted by the maxima of the max(DOS) curves defined in a manner similar to Fig.~\ref{extractMagicAngles_0_001}.
(right) The estimated $\omega_{AB^{\prime}}$(solid) and $\omega_{AA^{\prime}}$(dashed) by Eq.~(\ref{Eq:omega}) as a 
function of twist angle $\theta$. We observe that the main contribution to the predictions from the fully-relaxed calculations comes from the in-plane relaxation where the corresponding lattice reconstruction becomes weaker with increasing twist angle hence providing grounds for the reduction of the magic angle with respect to the analytical predictions based on the constant tunneling from the reference t2G system, as observed in panel \textbf{a.}.
}
\label{analyticVSNumeric}
\end{center}
\end{figure}

Here we find the numerical magic angle predictions of realistic lattice relaxed t$N$G systems  
that include both intra- and inter sublattice interlayer tunneling and using tight-binding calculations. 
Due to computational limitations when performing exact diagonalization to obtain the electronic band structure we use Lanczos recursion methods that allows to efficiently obtain the DOS on arbitrarily large systems with tens of million atoms which allows us to probe a fine grid of commensurate angles, linking the magic-angle flat bands to maxima in the DOS as a function of twist angle. 
We specifically focus on the first magic angle of t$N$G denoted as $\theta^{(N)}_{1}$ that 
from here onwards we refer to as simply magic angle. 
We have scanned the relaxed geometries at steps of approximately $\Delta \theta \sim 0.005^{\circ}$. 
The real space Hamiltonian includes both intra and inter sublattice interlayer tunneling as well as the lattice relaxation effects unlike the analytical model that only has inter-sublattice tunneling for rigid lattices.
%
%
The DOS peaks show a well defined maximum when we allow a sufficiently large broadening $\eta \sim 3.5$~meV, which is smaller but comparable to the bandwidths of the low energy nearly flat bands of typically $\sim 20$~meV, see Fig.~\ref{t3G_DOS_EBS} for t3G in the Appendix for a comparison between the electronic band structure and the corresponding features in the DOS using exact diagonalization. In our Lanczos calculations we have used two different broadening values to identify the magic angles. The width of the DOS peaks indicates the uncertainty in the magic twist angle values where experimental strong correlations are expected. Using a smaller $\eta \sim 0.175$~meV we can resolve finer undercover peaks within this broader peak that reflects the fine band structure features buried within a single nearly flat band. This behavior allows to define a {\em range of magic twist angles} centered around a value rather than a sharply defined unique magic angle value as we commonly understand.
This is illustrated respectively for large and smaller broadening 
in Figs.~\ref{extractMagicAngles_0_001} 
of the Appendix for t2G up to t20G and AT-graphite expanding on earlier discussions on t2G presented in Ref.~\cite{1910.12805}.
In Table.~\ref{tab:magicangle} and Fig.~\ref{analyticVSNumeric} we compare the magic angles obtained from the real space numerical calculations against a t$N$G continuum model with fixed interlayer tunneling as parametrized in Ref.~\cite{chebrolu2019}. 
\begin{table}[tbhp]
\begin{tabular}{lllll}
\hline
$N$ & Real-space & Continuum & Ref.~\cite{khalaf2019}  & Ref.~\cite{harvard2021}\\
 & relaxed  &~\cite{chebrolu2019}~&&\\
\hline
2 & 1.080   & 1.06 & 1.08   & 1.09  \\
3 & 1.510   & 1.49 & 1.527  & 1.49  \\
4 & 1.730   & 1.70 & 1.748  & 1.68  \\
5 & 1.845   & 1.84 & 1.871  & 1.79  \\
6 & 1.895   & 1.91 & 1.946  &       \\
10& 2.035   & 2.03 & 2.07   &       \\
20& 2.020   & 2.08 & 2.136  &       \\
Bulk& 1.90 & 2.12 & 2.16   &       \\
\hline
\end{tabular}
\caption{
Supporting numerical data for Fig.~\ref{analyticVSNumeric}. In the real-space relaxed calculations in the second column obtained using $\eta=3.5$~meV we see a reduction of the magic angle value $\theta$ with respect to the continuum model values with fixed interlayer tunneling in the third column as we progressively increase the number of layers $N$ and hence the respective first magic angles. We include as reference the predictions of the analytical values~\cite{khalaf2019} 
and continuum relaxed models~\cite{harvard2021} in columns four and five.
}
\label{tab:magicangle}
\end{table}
Provided that the interlayer tunneling model remains the same, the magic angle 
$\theta^{\rm bulk}_{1} = 2 \theta^{(2)}_{1}$ in the limit of infinite $N$ bulk AT-graphite systems 
is expected to be twice the value for t2G, 
namely $\theta^{(2)}_1 = 1.08^{\circ}$, since we have two interacting interfaces per unit cell 
as noted in Eq.~(\ref{Eq:continuumHamilbulk}).
The numerical magic angles resulting from lattice relaxed calculations 
using a broadening of $\eta=3.5$~meV in Eq.~(\ref{eqDos}) shows smaller 
magic angles such that $\theta^{\rm bulk}_{1} < 2 \theta^{(2)}_{1}$, 
with the large $N=20$ magic angle going as low as $\theta^{(20)}_1 = 2.02^\circ$ and the AT-graphite limit dipping all the way down to $\theta^{\rm bulk}_{1} = 1.9^\circ$. Similarly, a wider range of magic angles can be identified in the DOS peaks
when we use the smaller $\eta=0.175$~meV broadening. 

We observe a gradual decrease of the numerical magic angle with respect to the fixed interlayer coupling continuum model with increasing number that can be understood from an effective reduction of interlayer tunneling strength we discuss in the following. 
%
%
The evolution trend of the magic angles with lattice relaxations can be understood paying attention to its relationship with the interlayer tunneling in few layers systems, particularly t2G, keeping in mind that 
the magic angle will increase when the integrated interlayer tunneling $\omega$ between unequal sublattice sites between contiguous layers becomes larger. 
In Fig.~\ref{analyticVSNumeric} {\bf b.} we show the
comparison of the magic angles obtained for rigid systems with a fixed interlayer equilibrium distance of $c_0 = 3.313$~\AA, 
corresponding to AB or BA stacking interlayer distances of untwisted bilayer systems using our EXX-RPA force fields, the in-plane-only relaxed with the same $c_0$, the out-of-plane-only relaxed, and the fully relaxed geometries for t2G
that reveal the different elements entering the interlayer tunneling that affect the magic angle values. 
Because the interlayer distance of the rigid model (blue) is set to the shortest local interlayer equilibrium distance that maximizes tunneling, it is natural to expect that the predicted magic angle is larger than the out-of-plane relaxed (green) case that separates further the average interlayer distance between layers. 
The in-plane relaxed (orange) case that keeps the same interlayer distance as the rigid systems shows the largest magic angles. This behavior indicates that the surface-integrated inter-sublattice interlayer tunneling $\omega$ is effectively enhanced thanks to the increase of AB and BA local stacking areas when in-plane strains are allowed. 
Then a full relaxation that includes the out-of-plane corrugations will increase the average interlayer distance, reducing in turn the interlayer tunneling and therefore the magic angle. %
Using an electronic and atomic structure model calibrated to 
have a magic angle of $1.08^{\circ}$~\cite{1910.12805} for the fully relaxed system  
we observe that the rigid t2G's magic angle is 1.05$^{\circ}$, 
the in-plane-relaxation enhances it to $1.09^{\circ}$, and an
out-of-plane relaxation reduces it down to 1.02$^{\circ}$.
Likewise for t3G we see a similar trend of $1.47^{\circ}$, $1.52^{\circ}$ and $1.44^{\circ}$ respectively for fully relaxed, in-plane only and out-of-plane only. 
We can thus conclude that the drop in magic angle with respect to the continuum fixed interlayer tunneling model of the lattice relaxed t$N$G model for increasing $N$ is mainly because of the progressive decrease of in-plane strains for larger $\theta$, 
and therefore of interlayer tunneling $\omega$ between different sublattice atoms in contiguous layers.

We note that in the bulk limit we have an additional ingredient that enhances the drop in the magic twist angle value to $1.9^\circ$ because the periodicity along the $z$-axis implies that all layers are modeled to be flat and are free of corrugation. 
In this limit the equilibrium interlayer distance predicted by our EXX-RPA informed force field on the $2.00^\circ$ twist angle used to perform the calculations has a constant interlayer distance value of $3.328$~\AA\ that is slightly larger than the AB-stacking value of $3.313$~\AA\ of untwisted bilayers used for the rigid calculations of our t$N$G models. This larger interlayer distance of rigid AT-graphite reduces the inter-layer coupling strength and therefore results in a decrease of the predicted magic angle when compared to finite multilayer systems.

To further clarify the evolution of interlayer separation when transitioning from finite t$N$G systems to the AT-graphite bulk we present the average interlayer distance as a function of twist angle $\theta$ and number of layers $N$ in the Appendix. Intuitively we can understand this behavior from the fact that the AA stacking regions, that have the farthest local equilibrium interlayer separation, increases with twist angle. On the other hand, the corrugations are expected to gradually decrease for increasing number of $N$. 
In terms of the angle-dependence in Fig.~\ref{bulkEqDist}, we observe that the average and constant bulk interlayer distances increase with twist angle. Using the recipe to assign a specific local stacking to each atom in a twisted relaxed system introduced in Ref.~\cite{1910.12805}, and using the respective area ratio of these stackings, we can predict the bulk interlayer distance trend using Eq.~(\ref{ratioEq}). We finally observe that the free-standing t2G counterpart produces a similar increasing trend with twist angle but with slightly larger interlayer distances thanks to the presence of out-of-plane corrugations.
We show in Fig.~\ref{averageDistWithN}, for $\theta=2^\circ$, 
the $N$-dependence of a tNG system that converges monotonically 
to the bulk inter-layer distance when we use the simplifying constraint of keeping 
the outer-layer atoms in the same $xy$ plane to remove the corrugations. 
Without this constraint, in the fully relaxed system, the average interlayer 
distances converge more slowly to the bulk behavior as a function of $N$, 
showing a local minimum at around $N=14$. 
This behavior is in qualitative agreement with the maximum seen for $N=10$ in Fig.~\ref{analyticVSNumeric}.

\subsection{Electronic band structures of t$N$G multilayers and Hofstadter butterflies}

\begin{figure*}[htbp]
\begin{center}
\includegraphics[width=1\textwidth]{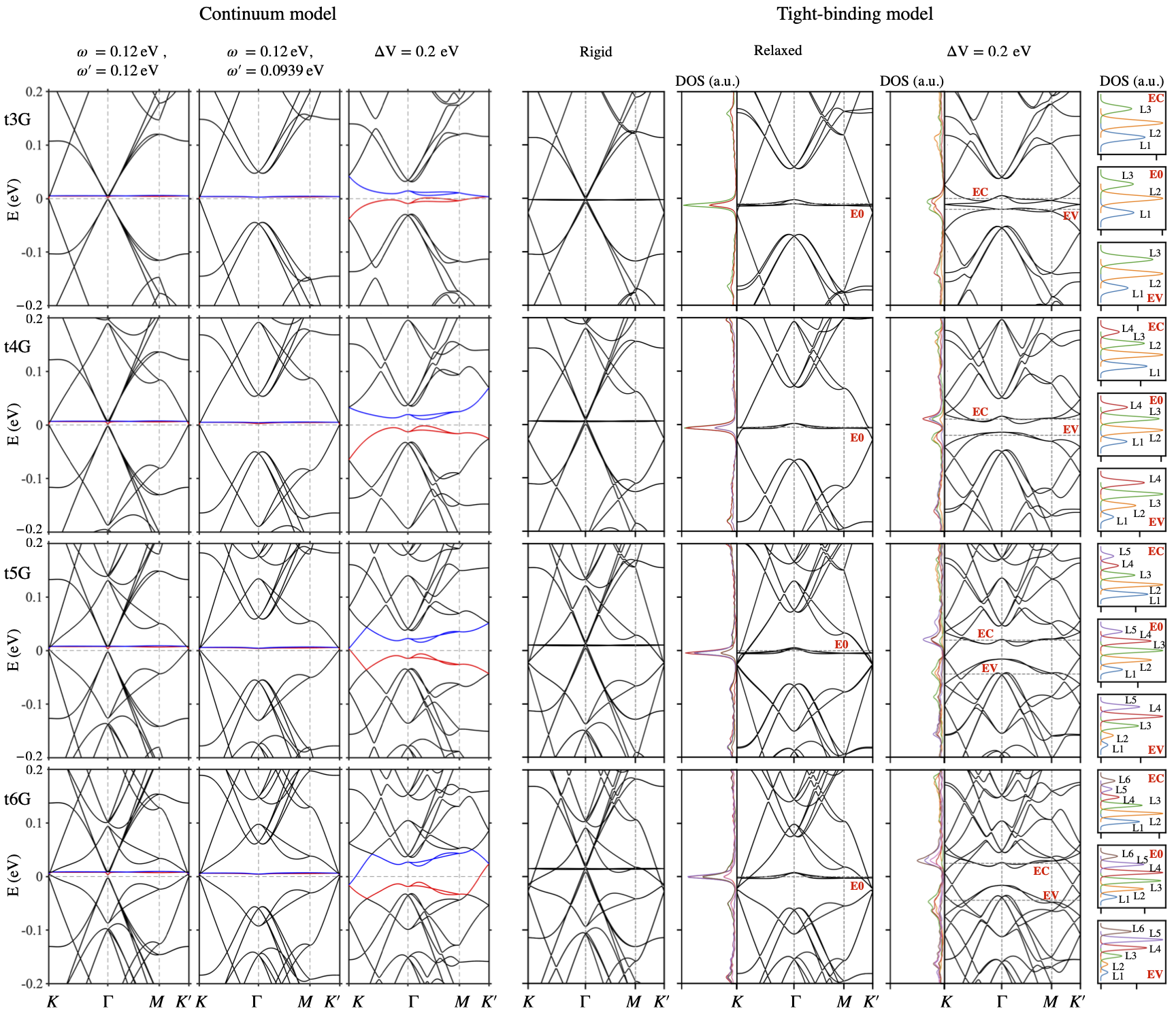}
\caption{ Electronic band structure of t$N$G for $N=3$ up to $N=6$ for the magic angles based on the criterion of minimizing the bandwidth at the $\Gamma-$point leading to values of $\theta_\text{eff} = 1.51^\circ$, $1.72^\circ$, $1.84^\circ$ and $1.9^\circ$ respectively,
 where we illustrate the lattice relaxation and perpendicular electric field effects. These values are only slightly different from the experimentally more relevant magic angle predictions $\theta$ based on the criterion of maximizing the DOS from Fig.~\ref{analyticVSNumeric} and Table~\ref{tab:magicangle} and these differences have a negligible impact on the conclusions drawn from this figure. The physical commensurate angles $\theta_\text{cell}$ and $S^{\prime}$ parameters from Eq.~\ref{effectiveS} that are used to simulate current effective angles $\theta_\text{eff}$ are listed in Table.~\ref{tab:systemparameters}. 
{\em Left panel:} Continuum model electronic bands at the $K$-valley. 
The left most column shows the equal tunneling $ \omega' = \omega = 0.12$~eV rigid model bands and distinct $ \omega' = 0.0939$~eV, $\omega = 0.12$~eV tunneling that partially 
accounts for out of plane lattice relaxation effects. 
The interlayer potential modified bands on top of this relaxed model illustrated in the third column
shows the splitting into $N$ Dirac cones manifest at the $K$ and 
$K^{\prime}$ points in the moire Brillouin zone. 
{\em Right panel:}
The electronic band structures for the rigid and relaxed systems at twist angles with maximum flatness of the central bands corresponding to bandwidth minima at the $\Gamma$-point. Contrary to the continuum bands, the lattice calculations show simultaneously both valleys $K$ and $K^{\prime}$. 
The layered-resolved DOS indicates that the states corresponding to the flat bands are located preferentially at the central layers of the system. 
Comparison between rigid and relaxed structures illustrates the overall electron-hole symmetry breaking introduced by the relaxation effects and opening of band gaps at the $\Gamma$ point. The interlayer potential difference $\Delta V = 0.2$~eV between top and bottom layers, see Eq.~(\ref{electricEq}), 
that can be introduced by a perpendicular electric field
tends to broaden the bandwidth of the low energy bands. 
The 4-layers system uniquely shows a bandgap in the electronic band structure, and the layer-resolved DOS show that the electric-field induced degeneracy-lifted low energy bands are leading to layer resolved charge polarization where we label the layers from bottom to top from L1 to L6 as defined in Fig.~\ref{energetics}. 
} 
\label{relaxationEffect}
\end{center}
\end{figure*}

%
In the following we discuss the effects of lattice relaxation in the
electronic band structures of t$N$G systems. 
For representation convenience the band structures are plotted in the Brillouin zones 
of smallest superlattice geometries whose periods are coincident with those of moire patterns. 
Because this type of coincidence only happens at discrete twist angle sets that are separated from each other we update the original interlayer tunneling prefactor $S$ in Eq.~(\ref{STCtunneling}) to $S^{\prime}$ of Eq.~(\ref{effectiveS}) to calculate the effective twisted angle $\theta_{\rm eff}$ given in Eq.~(\ref{effTheta}). 
The criterion to define the flattest band is based on the minimization of the band width at the 
$\Gamma$-point, leading to slightly different values than the experimentally more relevant ones predicted from DOS calculations in Sect.~\ref{DOSSect}. 
We thus refer to the magic angles represented in the electronic band structure calculations 
as $\theta_{\rm eff}$ and the ones from the DOS calculations as simply $\theta$.
The band structures of t3G up to t6G are shown in Fig.~\ref{relaxationEffect}
both for the continuum model with fixed interlayer interactions and lattice-relaxed real space tight-binding calculations, the latter with and without perpendicular electric fields. 
The relaxation separates the high energy bands at $\Gamma$ from the low energy bands~\cite{nam2017} 
and introduces an electron-hole asymmetry that becomes more pronounced in the tight-binding calculations. 
%
%
%
%
%
%
%
%
%
%
The layer-resolved density of states for each of these systems show that the flat band states are primarily populated by electrons associated with the atoms in the inner layers.
When applying an electric field to these systems in Fig.~\ref{relaxationEffect}, we notice that the layer-resolved DOS in the side-panels induces an electron-hole asymmetry in the degeneracy lifted flat band where the electron states are mostly located at the bottom layers while the hole states are being pushed towards the top layers. The corresponding band structures confirm the expected hybridization between the Dirac bands and the flat bands and, quite notably, show a clear real gap in the case of the t4G system~\cite{harvard2021}. The t5G system sees a robust crossing of the t4G gap by a Dirac cone at the K-point, while the t6G system sees two such crossing bands.
\begin{figure*}[htbp]
\begin{center}
\includegraphics[width=1.0\textwidth]{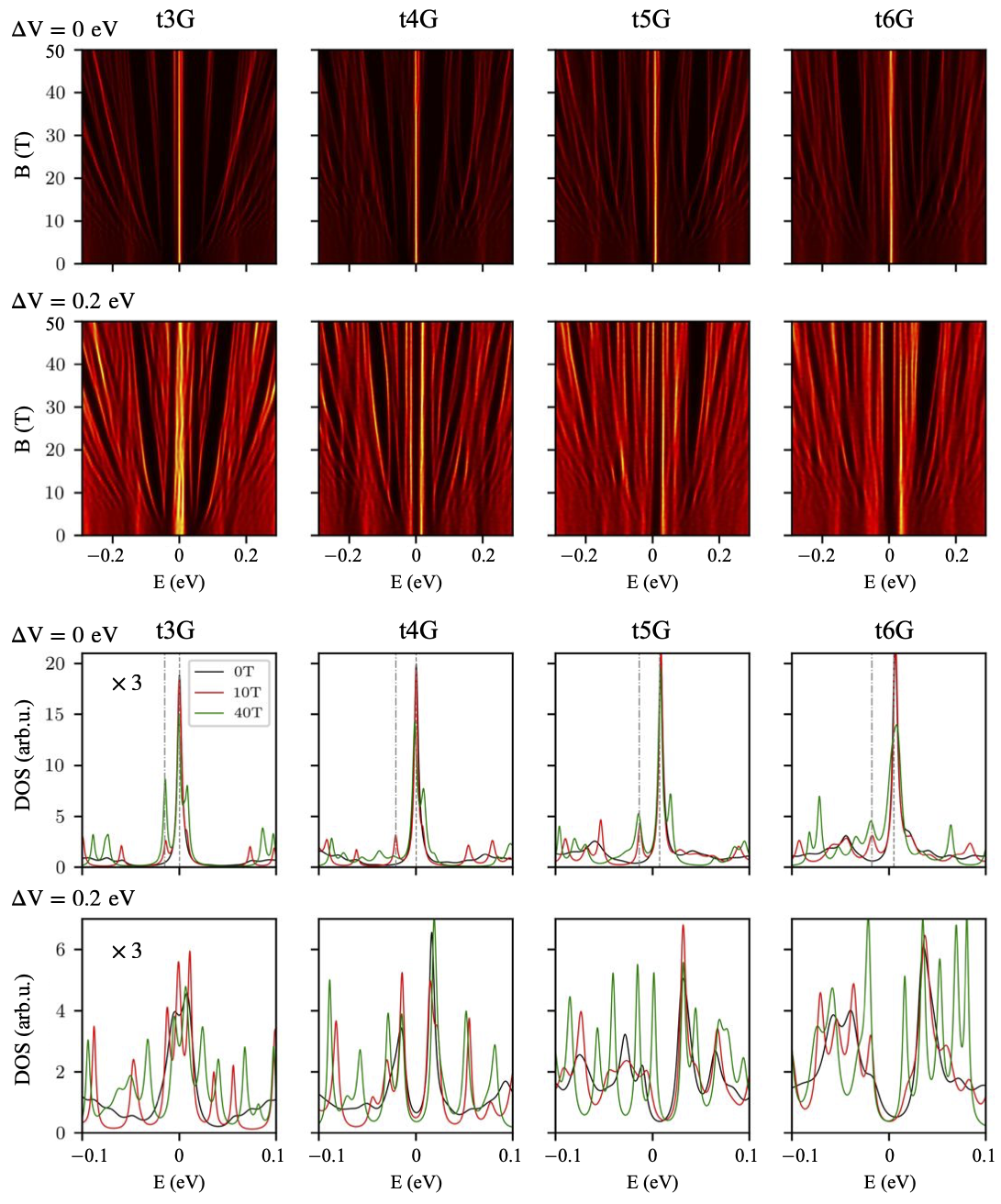}
\caption{(color online) Hofstadter Landau level maps based on the DOS (top) and DOS curves at selected magnetic fields (bottom) for t$N$G with $N=3,4,5,6$ with and without $\Delta V = 0.2$~eV top-bottom potential energy difference as defined in Eq.~(\ref{electricEq}). The angles $\theta_\text{eff}$ are the same ones as in Fig.~\ref{relaxationEffect} for which the physical commensurate angles and $S^{\prime}$ parameters are listed in Table.~\ref{tab:systemparameters}.
The amplitude of the t3G DOS curves have been multiplied by three to match the scale of the $N>3$ systems that have higher flat band peaks. 
The $N=3$ system doesn't show signs of a band gap in the presence of an interlayer potential difference 
while for the other $N$ considered the electric field-induced band gap due to DOS suppression 
in the quantum Hall regime increases with number of layers. 
The dashed vertical lines in the $\Delta V =0$ DOS plots indicate the flat band energies 
and the zeroth Landau Level corresponding to the lowest energy Dirac cone near charge neutrality. 
}
\label{magneticField}
\end{center}
\end{figure*}
\begin{figure*}[htbp]
\begin{center}
\includegraphics[width=1.0\textwidth]{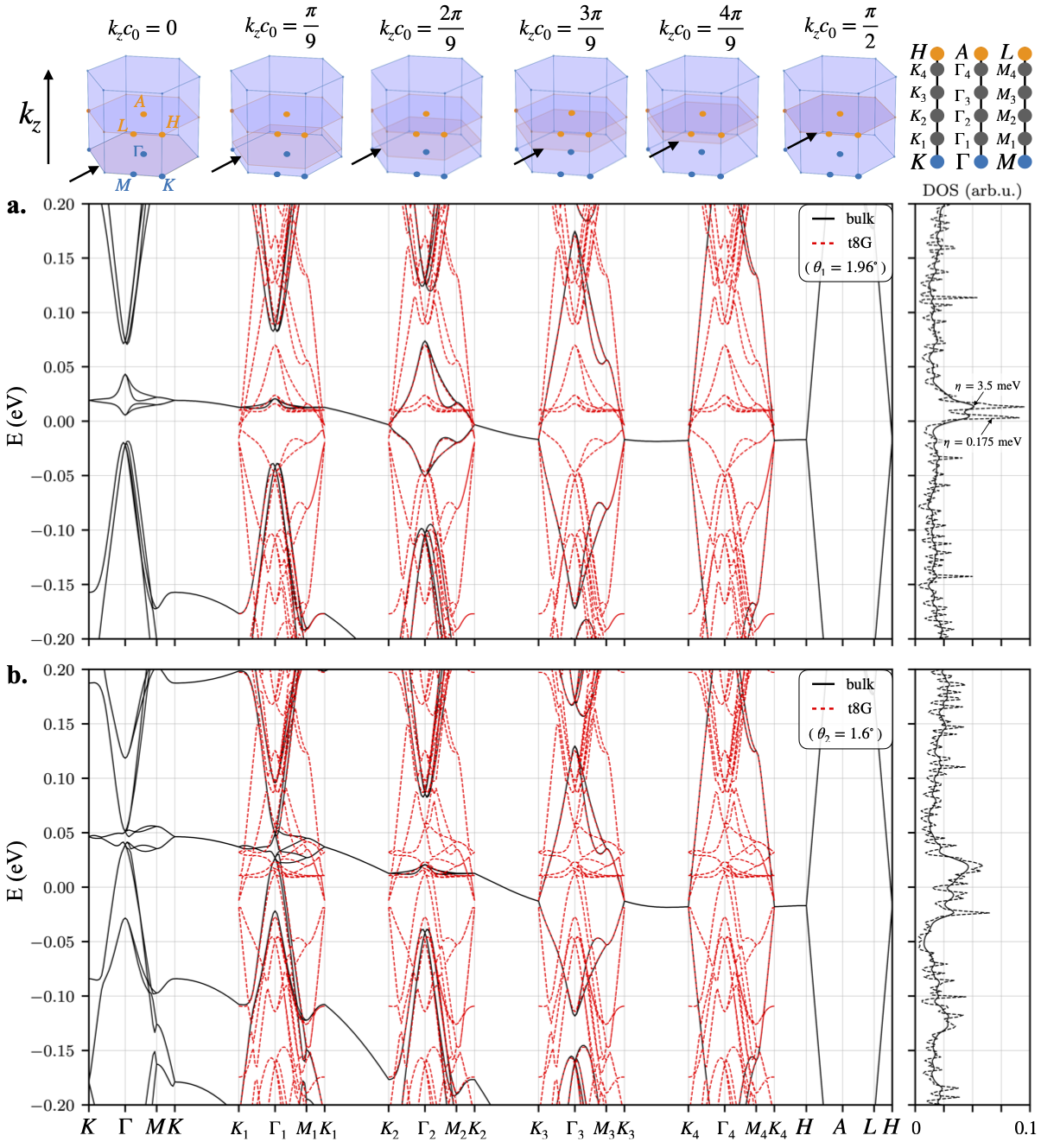}
\caption{(color online)
Comparison of bulk AT-graphite electronic bands in black at select $k_z c_0$ values given in Eq.~(\ref{kzrel}) corresponding to $[K_i, \Gamma_i, M_i, K_i]$ planes against lattice relaxed t8G bands in red.
Top {\bf a.} and bottom {\bf b.} panels represent two effective twist angles 
$\theta_{\rm eff} = 1.96^{\circ}$ and $\theta_{\rm eff} = 1.6^{\circ}$
corresponding to the magic angle of the first and second layers
and actual calculations were done in a commensurate cell with $\theta = 2.00^{\circ}$ using 
$S^\prime = 0.912$ and $1.119$.
The $[K, \Gamma, M, K]$, and $[H, A, L, H]$ planes corresponding to $k_z c_0 = 0, \pi/2$ can be mapped by a finite system for $N \rightarrow \infty$. 
In the bottom panel {\bf b.}, we see that the second bilayer band of t8G becomes very flat in the $[K_2, \Gamma_2, M_2, K_2]$ plane. These panels corresponding to different effective angles illustrate the coexistence of a continuum of magic angles in the bulk geometry. 
We provide in the Appendix Fig.~\ref{continuumBulk} a similar comparison based
on continuum models without including lattice relaxations.}
\label{BulkEBS}
\end{center}
\end{figure*}
We further illustrate the gap opening predicted for the systems with $N \leq 6$ by calculating the Landau level density of states diagrams up to $50$~T for each of the systems in Fig.~\ref{magneticField}. We indicate with dashed and dashed-dotted lines the location of the flat band and zeroth Landau level of (one of the) Dirac cones, respectively. The well-resolved separation and the suppressed density of states regions reveals that the flat bands are energy-pinned in the quantum Hall regime in $N>3$ systems. This effect can be verified when we apply electric fields that can decouple the effective Dirac cones and flat band contributions as observed in the band structures of Fig.~\ref{relaxationEffect}.
%
%

%
%
\subsection{From finite t$N$G multilayers to bulk AT-graphite bands}
\label{bulkSection}
Here, we discuss the electronic structure of t$N$G finite stacks in relation to bulk 
AT-graphite bands in the limit of $N \rightarrow \infty$.
Specifically, we verify the mapping of t$N$G bands onto those of bulk AT-graphite with 
different $k_z$.
This mapping can be achieved by relating the eigenvalues of a nearest neighbor linear 1D-chain of layers in a t$N$G system with the interlayer tunneling values of effective t2G systems it is made of. 
The preference of charge accumulation in the middle layers for the magic angle bands shown in Fig.~\ref{relaxationEffect} follows the wave function structure of the largest eigenvalue eigenstate of this 1D-linear chain, and therefore outer layer charge accumulation will be favored for smaller magic angles corresponding to reduced tunneling effective twisted bilayers.
We start by briefly discussing the decomposition of t$N$G bands into a sum of 
twisted bilayer models at different twist angles, plus a single layer for odd number of layers, 
that were obtained through singular value decomposition of the layers hopping matrix
where odd and even layers were distinguished in Ref.~\cite{khalaf2019}.
It turns out that this decomposition follows from the same type of basis change used to express a Bernal 
stacked $N$-multilayer graphene in terms of $[N/2]$ effective bilayers with renormalized 
interlayer tunneling values, plus a single layer graphene for odd $N$~\cite{koshino2008}.
We denote as $\psi_{k}( \ell )$ the layer resolved wave function amplitudes of the eigenstates
$\left| \psi_{k} \right>$ where $\ell = 1, 2, \hdots, N$ are the layer indices. 
The t$N$G multilayer Hamiltonian can be decomposed into smaller subsystems 
by choosing an appropriate basis of odd layer symmetric or even layer antisymmetric wave functions
based on the eigenstates of opposite sign eigenvalues  $\lambda_{k} = - \lambda_{N + 1 - k}$
\begin{equation}
    \begin{aligned}
    |\tilde{\psi}_{k}^{\rm odd} \rangle 
    &= \frac{ | \psi_{k }  \rangle + | \psi_{N + 1 -k } \rangle }{\sqrt{2}}, \\ \, \quad
      |\tilde{\psi}_{k}^{\rm even}\rangle &= \frac{ | \psi_{k }  \rangle - | \psi_{N + 1 -k } \rangle }{\sqrt{2}}.
    \label{basis}
    \end{aligned}
\end{equation}
with $k=1,2,\cdots,N_e$. When $N$ is odd and $k=N_o$ we have for the zero eigenvalue state $|\tilde{\psi}_{k}^{\rm odd}> = | \psi_{k }  \rangle$.
The eigenstates corresponding to the $k^{\rm th}$ eigenvalue following Eq.~(\ref{eigenstates}) and the associated symmetric/antisymmetric forms are shown in Fig.~\ref{sketch} in the appendix. This figure shows the preferential accumulation of charge in the middle layers for $k = 1$ and how the charge progressively redistributes towards the outer layers for larger $k$, or equivalently for the magic angles of the second or third effective bilayers.

\begin{figure}[tbhp]
\begin{center}
\includegraphics[width=1.0\columnwidth]{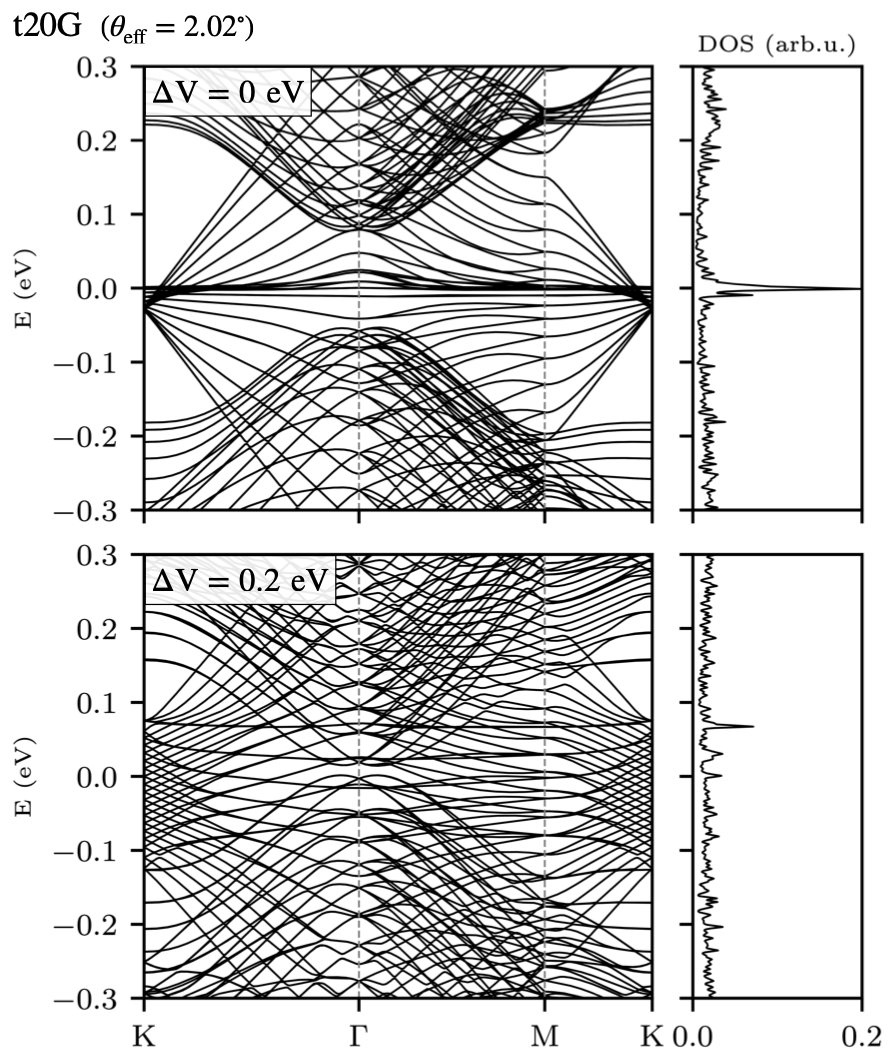}
\caption{(color online) 
Electronic band structure and total DOS for a 20-layer twisted multilayer system for the effective first magic angle at $\theta_{\rm eff} = 2.02^{\circ}$ corresponding to a minimal width at the $\Gamma-$point (equal to the $\theta=2.02^\circ$ value based on the criterion of maximizing the DOS, 
for zero (top) and $0.2$~eV top-bottom interlayer potential energy difference (bottom). 
The interlayer potentials are modeled to linearly vary throughout the layers following Eq.~(\ref{electricEq}). The physical commensurate angle and associated $S^{\prime}$ factor are summarized in Table~\ref{tab:systemparameters}.
We can clearly resolve the 10 sets of bilayer systems and we notice that the electron and valence bands at the $\Gamma$-point overlap with the flat bands in presence of an electric field.
}
\label{twentyLayersElectricField}
\end{center}
\end{figure}

If we represent the Hamiltonian given in Eq.~(\ref{Eq:continuumHamil}) in this modified basis that distinguishes odd and even layers we get
\begin{equation}
\tilde{{\cal{H}}}^{\rm (dec)}_{\bm{k}} \\
= 
\begin{pmatrix}
h_{oo}^{AA}\bm{1}_{o}  &h_{oo}^{AB}\bm{1}_o    &T_{oe}^{AA}\bm{\Lambda} &T_{oe}^{AB}\bm{\Lambda}\\
h_{oo}^{BA}\bm{1}_{o}  &h_{oo}^{BB}\bm{1}_o    &T_{oe}^{BA}\bm{\Lambda} &T_{oe}^{BB}\bm{\Lambda}\\
T_{eo}^{AA}\bm{\Lambda}&T_{eo}^{AB}\bm{\Lambda}&h_{ee}^{AA}\bm{1}_e     &h_{ee}^{AB}\bm{1}_e\\
T_{eo}^{BA}\bm{\Lambda}&T_{eo}^{BB}\bm{\Lambda}& h_{ee}^{BA}\bm{1}_e    &h_{ee}^{BB}\bm{1}_e
\end{pmatrix}
.
\end{equation}
We can further transform the matrix into effective t2G-like Hamiltonian diagonal blocks with 
tunneling strengths proportional to the eigenvalue $\lambda_k$ by further grouping the basis into alternating sublattices for different layers
\begin{equation}
\begin{aligned}
{{\cal{H}}}_{\bm{k}}^{\rm (dec)}
=&P \tilde{{\cal{H}}}_{\bm{k}} P^{\dagger} \\ 
=&{\cal{H}}_{\bm{k} \lambda_{1}}^{(t2G)}\oplus {\cal{H}}_{\bm{k} \lambda_{2}}^{(t2G)}\oplus \cdots \oplus {\cal{H}}_{\bm{k}\lambda_{N_e}}^{(t2G)}\\ 
 &\left(\oplus {\cal{H}}_{\bm{k}}^{(1G)}\right)
\end{aligned}
\end{equation}
as discussed in Refs.~\cite{koshino2008,khalaf2019}.
Here $P$ is the transformation matrix that takes the sublattice $ s = A, B$ labeled
basis $\left( \psi_{A_1}, \psi_{B_1}, \psi_{A_2}, \psi_{B_2},\cdots, \psi_{A_N}, \psi_{B_N}\right)^T$ to 
$\left( \psi_{A}^{\rm o}, \psi_{B}^{\rm o}, \psi_{A}^{\rm e}, \psi_{B}^{\rm e}\right)^T$ where
$\psi_{s}^{\rm o} = ( \psi_{s_1}, \psi_{s_3},\cdots,\psi_{s_{2N_o-1}} )^T$ and $\psi_{s}^{\rm e} = ( \psi_{s_2}, \psi_{s_4},\cdots,\psi_{s_{2N_e}} )^T$ 
where $N_{o/e}$ is the number of odd/even layers. The
$\Lambda = \sum_k \ \lambda_{k} \delta_{k,n}$ are diagonal $N_o \times N_e$ matrices with $n,k=1,2,\cdots, N_e$.
A detailed derivation of the decomposition procedure following the basis change is presented in~\ref{derivationAppendix}.
%
From Eq.~(\ref{Eq:continuumHamilbulk}) the bands of bulk AT-graphite can be viewed as standing wave bilayer bands with renormalized interlayer tunneling strengths with periodic boundary conditions and therefore the few layers graphene bands can be related with those of bulk graphite at select $k_z$ by 
\begin{equation}
k_{z} c_0 = \frac{\pi k}{ N+1 }
\label{kzrel}
\end{equation} 
where we require the interlayer tunneling for bulk AT-graphite Eq.~(\ref{Eq:continuumHamilbulk})
and the effective bilayer Hamiltonian tunneling in Eq.~(\ref{1dchainEq}) to be the same.
To illustrate the t2G bands that exist within a t$N$G band structure in the Hamiltonian and the subsequent mapping 
to the AT-graphite we overlay in Fig.~\ref{BulkEBS} the bands for $N=8$ case on the AT-graphite bands at select 
$k_z$ values obtained at $\pi/9$, $2 \pi/9$, $3 \pi/9$ and $4 \pi/9$-$k_z c_0$ cuts.
In the first panel, we see a clear agreement between the bulk AT-graphite and the 
t8G bands for the first magic-angle of the first bilayer band where the bulk is corrugation-free by design and the free-standing finite system contains corrugation.
In the second panel, we set the effective twist angle to $\theta_{\rm eff} = 1.6^\circ$ 
by tuning the coupling strength $S^{\prime}$ to the first magic-angle of the 
second bilayer band at $k_z c_0 = 2\pi/9$-plane in the bulk band structure. In both cases we observe a good agreement between the bulk and the t8G system where the difference stems mainly from the different relaxation profiles.
We provide a similar bulk-finite mapping based on the continuum model in the Appendix in Fig.~\ref{continuumBulk} showing an even closer agreement due to the absence of relaxations.
%
Decoupling of the bands into effective bilayers with different Fermi velocities was illustrated for few layers t$N$G in Fig.~\ref{relaxationEffect} for $N=3, 4, 5, 6$
and we further illustrate this in Fig.~\ref{twentyLayersElectricField} for a t20G system at $\theta_{\rm eff} = 2.02^{\circ}$ twist.
We note that adding an electric field in t20G allows to resolve the Dirac cones of the associated effective bilayer bands and brings the higher energy conduction and valence bands closer to the charge neutral point.

\section{Summary and conclusions}
\label{conclusionsSection}
We have carried out the stacking dependent atomic structure calculations of t$N$G systems up to $N=6$ layers 
and shown that the highly symmetric AA$^\prime$AA$^\prime$\dots\-stacked sliding configuration is energetically the most stable of all possible alternatives indicating that these will be favored when devices are prepared experimentally. 
In particular, we find that there are barrier-free sliding paths towards the energetically stable symmetric stacking configuration. 

From an electronic structure perspective, the lattice relaxation leads to small reductions in the magic angle predictions compared to analytical or continuum model calculations with fixed interlayer coupling strengths. This behavior has been confirmed numerically from the $N=3$ all the way up to the bulk AT-graphite configuration. 
We have explained this behavior based on the progressive reduction of the integrated interlayer tunneling 
$\omega$ from small to large $N$ limit that is mainly due to the decrease of in-plane strains when the layers are twisted to have larger angles that in turn reduces the relative area of AB/BA stacking regions. 
The relaxed lattice geometries also lead to electron-hole asymmetry enhancement and isolation of the flat bands from the rest of the spectrum at the $\Gamma$-point in t$N$G systems. 

When we introduce a perpendicular electric field, charge is relocated from the inner layers towards the outer layers and leads to decoupled $N$ Dirac bands that acquire different Fermi velocity values. Only the $N=4$ layered system leads to gap opening in the presence of an electric field near the first magic angle, while one or two Dirac cones survive near the charge neutral point in the case of $N=5$ and $N=6$ systems. 

We have calculated the Landau levels in the quantum Hall regime 
for magnetic fields of up to $B=50$~T. Except for the t3G case, we have shown that sufficiently strong perpendicular electric fields can deplete the
density of states near charge neutrality for t$N$G systems 
leading to practically gapped phases. 

Finally, we showed that the band structures in multilayer t$N$G can be mapped closely onto 
bulk AT-graphite bands at select $k_z$ values following the decoupling of t$N$G bands 
into effective bilayer models where the interlayer tunneling values are proportional to the 
eigenenergies of a 1D-chain of layers, with small differences stemming from relaxation effects.
The eigenvector amplitudes in each layer indicate the charge densities associated to the flat 
bands where the largest eigenvalue eigenvectors tend the concentrate the charge at the middle layers while these redistribute towards the outer layers when we choose smaller twist angles.

\section*{Acknowledgement}
N.~L. was supported by the Korean National Research Foundation grant NRF-2020R1A2C3009142, 
Y.~J. Park was supported by grant NRF-2021R1A6A3A13045898 and A.~S. was supported by grant NRF-2020R1A5A1016518.
J.~A. was supported by the Korean Ministry of Land, Infrastructure and Transport (MOLIT) from 
the Innovative Talent Education Program for Smart Cities.
J.~J.  was supported by the Samsung Science and Technology Foundation under project SSTF-BAA1802-06.
We acknowledge computational support from KISTI through grant KSC-2021-CRE-0389, the resources of Urban Big data and AI Institute (UBAI) at the University of Seoul and the network support from KREONET.

\bigskip
\bibliography{all}

\providecommand{\newblock}{}
\begin{thebibliography}{10}
\expandafter\ifx\csname url\endcsname\relax
  \def\url#1{{\tt #1}}\fi
\expandafter\ifx\csname urlprefix\endcsname\relax\def\urlprefix{URL }\fi
\providecommand{\eprint}[2][]{\url{#2}}

\bibitem{cao2018}
Cao Y, Fatemi V, Demir A, Fang S, Tomarken S~L, Luo J~Y, Sanchez-Yamagishi J~D,
  Watanabe K, Taniguchi T, Kaxiras E, Ashoori R~C and Jarillo-Herrero P 2018
  {\em Nature\/} {\bf 556} 80 ISSN 1476-4687

\bibitem{Park2021}
Park J~M, Cao Y, Watanabe K, Taniguchi T and Jarillo-Herrero P 2021 {\em
  Nature\/} {\bf 590} 249--255
  \urlprefix\url{https://doi.org/10.1038/s41586-021-03192-0}

\bibitem{Hao2021}
Hao Z, Zimmerman A~M, Ledwith P, Khalaf E, Najafabadi D~H, Watanabe K,
  Taniguchi T, Vishwanath A and Kim P 2021 {\em Science\/} {\bf 371} 1133--1138
  \urlprefix\url{https://doi.org/10.1126/science.abg0399}

\bibitem{li2019}
Li X, Wu F and Das~Sarma S 2020 {\em Phys. Rev. B\/} {\bf 101}(24) 245436
  \urlprefix\url{https://link.aps.org/doi/10.1103/PhysRevB.101.245436}

\bibitem{Cea2019}
Cea T, Walet N~R and Guinea F 2019 {\em Nano Letters\/} {\bf 19} 8683--8689
  \urlprefix\url{https://doi.org/10.1021/acs.nanolett.9b03335}

\bibitem{PhysRevResearch.2.022010}
Wu F, Zhang R~X and Das~Sarma S 2020 {\em Phys. Rev. Research\/} {\bf 2}(2)
  022010
  \urlprefix\url{https://link.aps.org/doi/10.1103/PhysRevResearch.2.022010}

\bibitem{1901.09356}
Moriyama S, Morita Y, Komatsu K, Endo K, Iwasaki T, Nakaharai S, Noguchi Y,
  Wakayama Y, Watanabe E, Tsuya D, Watanabe K and Taniguchi T 2019 Observation
  of superconductivity in bilayer graphene/hexagonal boron nitride
  superlattices (\textit{Preprint} \eprint{arXiv:1901.09356})

\bibitem{Liu2020}
Liu X, Hao Z, Khalaf E, Lee J~Y, Ronen Y, Yoo H, Najafabadi D~H, Watanabe K,
  Taniguchi T, Vishwanath A and Kim P 2020 {\em Nature\/} {\bf 583} 221--225
  \urlprefix\url{https://doi.org/10.1038/s41586-020-2458-7}

\bibitem{2203.09188}
Nguyen V~H, Hoang T~X and Charlier J~C 2022 Electronic properties of twisted
  multilayer graphene (\textit{Preprint} \eprint{arXiv:2203.09188})

\bibitem{jeongminpark2021}
Park J~M, Cao Y, Watanabe K, Taniguchi T and Jarillo-Herrero P 2021 {\em
  Nature\/} {\bf 590} 249--255
  \urlprefix\url{https://doi.org/10.1038/s41586-021-03192-0}

\bibitem{jeongminpark2022}
Park J~M, Cao Y, Xia L, Sun S, Watanabe K, Taniguchi T and Jarillo-Herrero P
  2021 {\em arXiv:2112.10760\/}

\bibitem{PhysRevB.96.195431}
Leconte N, Jung J, Leb\`egue S and Gould T 2017 {\em Phys. Rev. B\/} {\bf
  96}(19) 195431
  \urlprefix\url{https://link.aps.org/doi/10.1103/PhysRevB.96.195431}

\bibitem{Bistritzer:2011ho}
Bistritzer R and MacDonald A~H 2011 {\em Proceedings of the National Academy of
  Sciences\/} {\bf 108} 12233--12237
  \urlprefix\url{https://www.pnas.org/doi/abs/10.1073/pnas.1108174108}

\bibitem{chebrolu2019}
Chebrolu N~R, Chittari B~L and Jung J 2019 {\em Physical Review B\/} {\bf 99}
  235417 ISSN 2469-9950

\bibitem{TRAMBLYDELAISSARDIERE:2013fa}
Trambly~de Laissardi\`ere G and Mayou D 2013 {\em Phys. Rev. Lett.\/} {\bf
  111}(14) 146601
  \urlprefix\url{https://link.aps.org/doi/10.1103/PhysRevLett.111.146601}

\bibitem{1910.12805}
Leconte N, Javvaji S, An J and Jung J 2019 Relaxation effects in twisted
  bilayer graphene: a multi-scale approach (\textit{Preprint}
  \eprint{arXiv:1910.12805})

\bibitem{jung2014}
Jung J, Raoux A, Qiao Z and MacDonald A~H 2014 {\em Phys. Rev. B\/} {\bf
  89}(20) 205414
  \urlprefix\url{https://link.aps.org/doi/10.1103/PhysRevB.89205414}

\bibitem{10.1103/physrevb.96.085442}
Jung J, Laksono E, DaSilva A~M, MacDonald A~H, Mucha-Kruczyński M and Adam S
  2017 {\em Phys. Rev. B\/} (\textit{Preprint} \eprint{1706.06016})

\bibitem{McKinnon1993}
McKinnon B and Choy T 1993 {\em Australian Journal of Physics\/} {\bf 46} 601
  \urlprefix\url{https://doi.org/10.1071/ph930601}

\bibitem{jung2013}
Jung J and MacDonald A~H 2013 {\em Physical Review B\/} {\bf 87} 195450 ISSN
  1098-0121

\bibitem{laissardire2012}
Laissardi\'ere G~T~d, Mayou D and Magaud L 2012 {\em Physical Review B\/} {\bf
  86} 125413 ISSN 1098-0121

\bibitem{Chittari2018}
Chittari B~L, Leconte N, Javvaji S and Jung J 2018 {\em Electronic Structure\/}
  {\bf 1} 015001

\bibitem{khalaf2019}
Khalaf E, Kruchkov A~J, Tarnopolsky G and Vishwanath A 2019 {\em Phys. Rev.
  B\/} {\bf 100}(8) 085109
  \urlprefix\url{https://link.aps.org/doi/10.1103/PhysRevB.100.085109}

\bibitem{lanczos}
Fan Z, Garcia J~H, Cummings A~W, Barrios-Vargas J~E, Panhans M, Harju A,
  Ortmann F and Roche S 2021 {\em Physics Reports\/} {\bf 903} 1--69 ISSN
  0370-1573 linear scaling quantum transport methodologies
  \urlprefix\url{https://www.sciencedirect.com/science/article/pii/S0370157320304245}

\bibitem{PhysRevB.103.045402}
Cresti A 2021 {\em Phys. Rev. B\/} {\bf 103}(4) 045402
  \urlprefix\url{https://link.aps.org/doi/10.1103/PhysRevB.103.045402}

\bibitem{Brenner_2002}
Brenner D~W, Shenderova O~A, Harrison J~A, Stuart S~J, Ni B and Sinnott S~B
  2002 {\em Journal of Physics: Condensed Matter\/} {\bf 14} 783--802
  \urlprefix\url{https://doi.org/10.1088/0953-8984/14/4/312}

\bibitem{PhysRevB.98.235404}
Wen M, Carr S, Fang S, Kaxiras E and Tadmor E~B 2018 {\em Phys. Rev. B\/} {\bf
  98}(23) 235404
  \urlprefix\url{https://link.aps.org/doi/10.1103/PhysRevB.98.235404}

\bibitem{Plimpton1995}
Plimpton S 1995 {\em Journal of Computational Physics\/} {\bf 117} 1--19
  \urlprefix\url{https://doi.org/10.1006/jcph.1995.1039}

\bibitem{PhysRevLett.97.170201}
Bitzek E, Koskinen P, G\"ahler F, Moseler M and Gumbsch P 2006 {\em Phys. Rev.
  Lett.\/} {\bf 97}(17) 170201
  \urlprefix\url{https://link.aps.org/doi/10.1103/PhysRevLett.97.170201}

\bibitem{Rowe2020}
Rowe P, Deringer V~L, Gasparotto P, Cs{\'{a}}nyi G and Michaelides A 2020 {\em
  The Journal of Chemical Physics\/} {\bf 153} 034702
  \urlprefix\url{https://doi.org/10.1063/5.0005084}

\bibitem{Carr2020}
Carr S, Li C, Zhu Z, Kaxiras E, Sachdev S and Kruchkov A 2020 {\em Nano
  Letters\/} {\bf 20} 3030--3038
  \urlprefix\url{https://doi.org/10.1021/acs.nanolett.9b04979}

\bibitem{harvard2021}
Ledwith P~J, Khalaf E, Zhu Z, Carr S, Kaxiras E and Vishwanath A 2021 {\em
  arXiv:2111.11060\/}

\bibitem{nam2017}
Nam N~N~T and Koshino M 2017 {\em Physical Review B\/} {\bf 96} 075311 ISSN
  2469-9950

\bibitem{koshino2008}
Koshino M and Ando T 2008 {\em Phys. Rev. B\/} {\bf 77}(11) 115313
  \urlprefix\url{https://link.aps.org/doi/10.1103/PhysRevB.77.115313}

\bibitem{PhysRevLett.120.046801}
Stauber T, Low T and G\'omez-Santos G 2018 {\em Phys. Rev. Lett.\/} {\bf
  120}(4) 046801
  \urlprefix\url{https://link.aps.org/doi/10.1103/PhysRevLett.120.046801}

\bibitem{Kim2016}
Kim C~J, S{\'{a}}nchez-Castillo A, Ziegler Z, Ogawa Y, Noguez C and Park J 2016
  {\em Nature Nanotechnology\/} {\bf 11} 520--524
  \urlprefix\url{https://doi.org/10.1038/nnano.2016.3}

\end{thebibliography}

\appendix

\section{Potential and elastic energy maps}

\begin{figure*}[htbp]
\begin{center}
\includegraphics[width=1.0\textwidth]{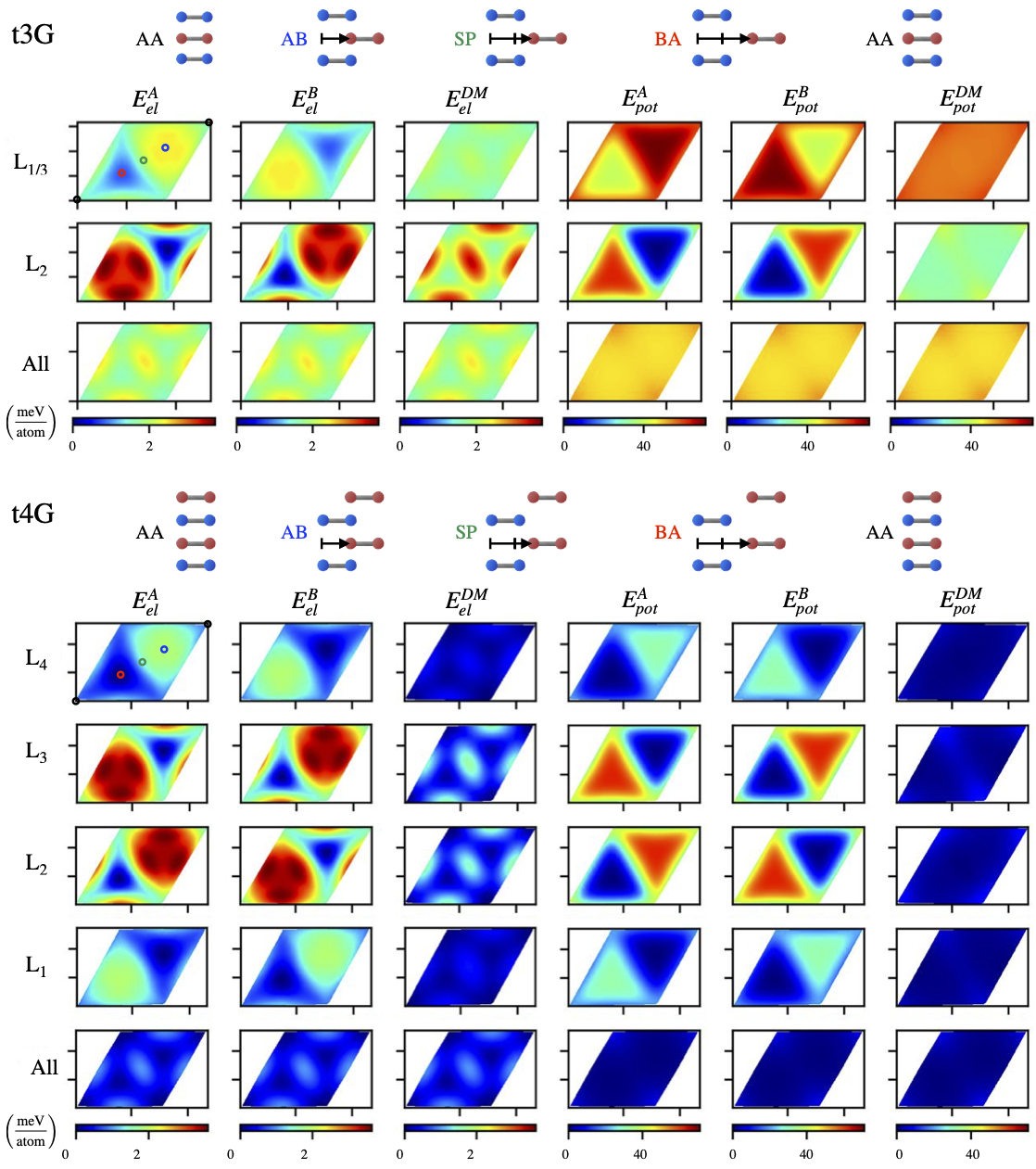}
\caption{
Layer-, sublattice-, and dimer (DM)-resolved maps of local elastic ($E_{\rm el}$) energy in Eq.~(\ref{elastic}) and potential ($E_{\rm pot}$) energy in Eq.~\ref{potential} for the minimum energy sliding t3G and t4G configurations are shown in the top and bottom panels respectively. Because the outer layers have only one interface the elastic and potential energy variations are roughly half of the inner layers that have two. 
We separate both the sublattice A and B contributions as well as the L$_1$ (equal to L$_3$) 
and L$_2$ layer contributions. DM refers to the dimer contribution which is the average of 
$A$ and $B$ sublattice contributions and \textit{All} is the average of all the different 
layer contributions. The energies in the figure are given in meV/atom. We apply a constant 
energy shift for both systems of $-7.397$~eV/atom for the elastic energies and $0.079$~eV/atom 
for the potential energies to ease visual comparison between the different panels. 
This brings to zero the smallest elastic or potential energy value shown in the graphs. 
We notice that the outer layers carry about half of the potential and elastic energy 
differences due to the single interface when compared with the middle layer(s) that have two. 
The average of \textit{all} layer contributions fluctuate around a constant average value by a relatively small amount with respect to variations in an individual layer. 
The potential energy contributions are greater in magnitude than the elastic energies 
and lead energetically favorable (AB/BA) and unfavorable regions (AA). 
The average elastic energy is highest at the $SP$ regions. Distinct layer/sublattice 
dependent elastic and potential energies reflect the chirality of the alternating 
twist angle signs.
}
\label{energetics3L}
\end{center}
\end{figure*}

In Fig.~\ref{energetics3L}, we provide the potential energy and elastic energy contributions for the t3G and t4G sytems. We separate these from our LAMMPS calculation by summing separately the pair-wise interactions between inter-layer atoms (potential energy) and intra-layer atoms (elastic energy). We map out the profiles using the elastic and potential energy contributions from each atom $i$ calculated as 
\begin{eqnarray}
E^{\rm el}_i &=& \sum_{j\in \text{layer i}} \phi_{ij}    \label{elastic} \\
E^{\rm pot}_i &=&  \sum_{j\notin \text{layer i}} \phi_{ij}  =  \sum_{j\in \text{ any layer}} \phi_{ij} - E^{\rm el}_i \label{potential}
\end{eqnarray}
where $\phi_{ij}$ are the pair-wise interactions between atoms $i$ and $j$ modeled by the DRIP potential~\cite{PhysRevB.98.235404} for interlayer interactions and the REBO2 potential~\cite{Brenner_2002} for intralayer interactions. 
The maps of the outer layers show a factor $\sim$1/2 in terms of the potential energy differences due to having only one interface with another layer.
As mentioned in the main text, based on the potential energy, the AA stacking shows the highest binding energy, or conversely the lowest free energy; the energetically favorable and unfavorable regions can thus be fully inferred from the potential energy maps. 
This stands in contrast with the layer resolved potential energy maps where AA-stacking corners do not appear as the energetically least favorable. 
In terms of the elastic energy, we observe that, when adding up all layers, the strongest strain is located at the center of the SP regions with the symmetry stacking AA and AB positions being the ones containing the least amount of elastic strain.
We further note that, due to the alternating rotation directions between consecutive layers, 
the layer and sublattice-resolved energy maps show alternating chirality both for their elastic and potential contributions. The chiral nature of t2G has been linked to strong optical activity at finite frequencies corresponding to transitions around the M point~\cite{PhysRevLett.120.046801, Kim2016}.
The matching trends for t3G and t4G suggest that similar conclusions can be drawn for increasing number of layers.

\section{Magic angle extraction and fine DOS features}
\label{magicAngleExtraction}
\begin{figure*}[htbp]
\begin{center}
\includegraphics[width=0.9\textwidth]{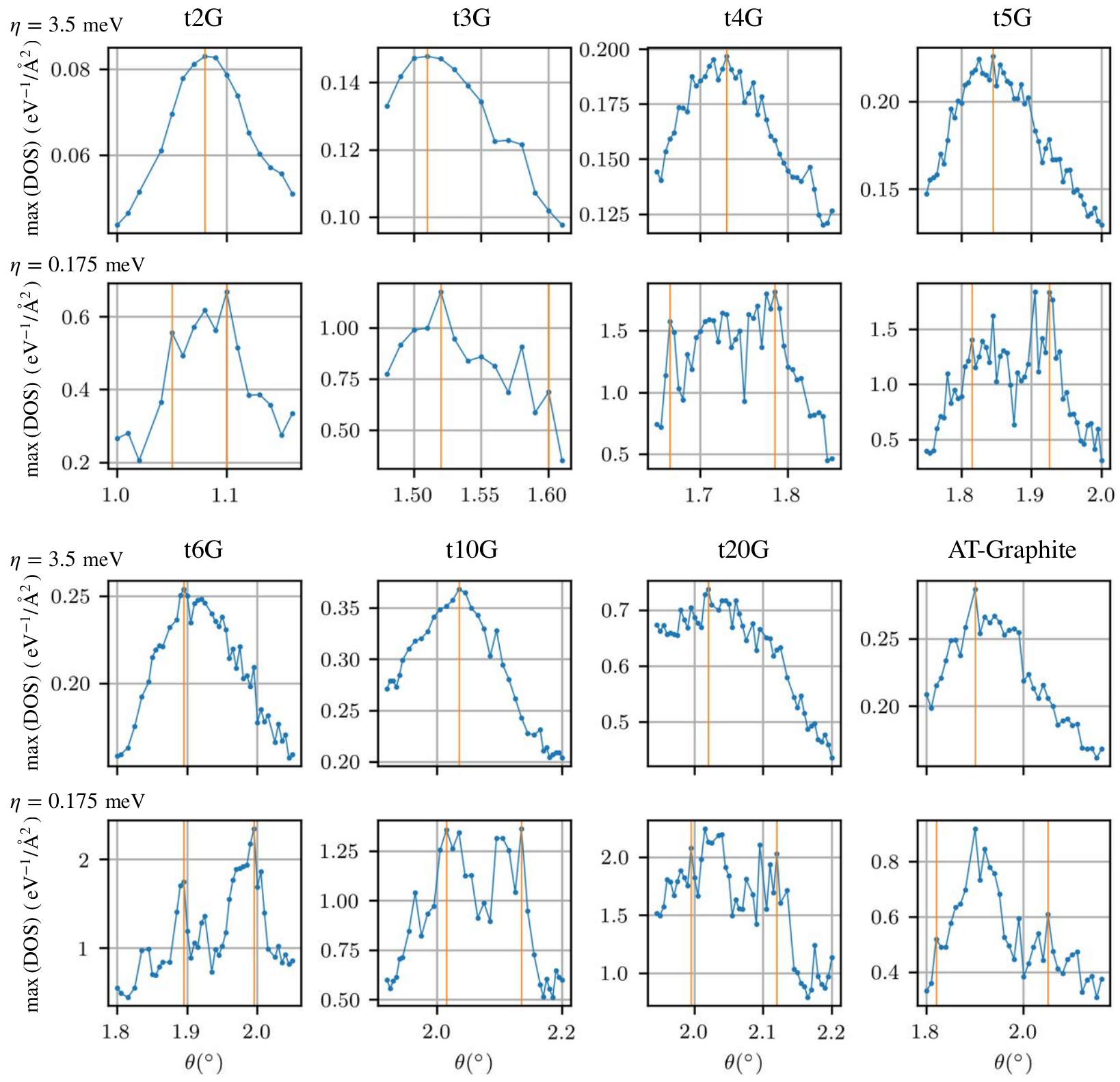}
\caption{(color online) Using a broadening $\eta$ of about (i) $3.5$ and (ii) $0.175$ meV in Eq.~(\ref{eqDos}), we extract the magic angles to be (i) the ones corresponding here to the highest value in the range plotted here where max(DOS) is the maximum of the DOS curve for a specific angle in a range close to the Fermi energy where the flat band is expected to be and (ii) the ones corresponding to the angles encompassed by the outer max(DOS) peaks that have an amplitude above an arbitrary threshold within the range plotted here. These values are then reported in Fig.~\ref{analyticVSNumeric}  and indicated here by orange lines. The finite resolution of angle points as well as the partially arbitrary choice in what is considered to be a peak causes some uncertainty in the extraction of these magic angles. We note that for simplicity the angle-sweep for the AT-Graphite system are done for a fixed interlayer distance of $3.328$~\AA, although a small angle-dependence of this distance is observed as illustrated in Fig.~\ref{bulkEqDist}. This simplification should affect the magic angle extraction by $0.02^\circ$ at most.
}
\label{extractMagicAngles_0_001}
\end{center}
\end{figure*}

\begin{figure}[tbhp]
\begin{center}
\includegraphics[width=1.0\columnwidth]{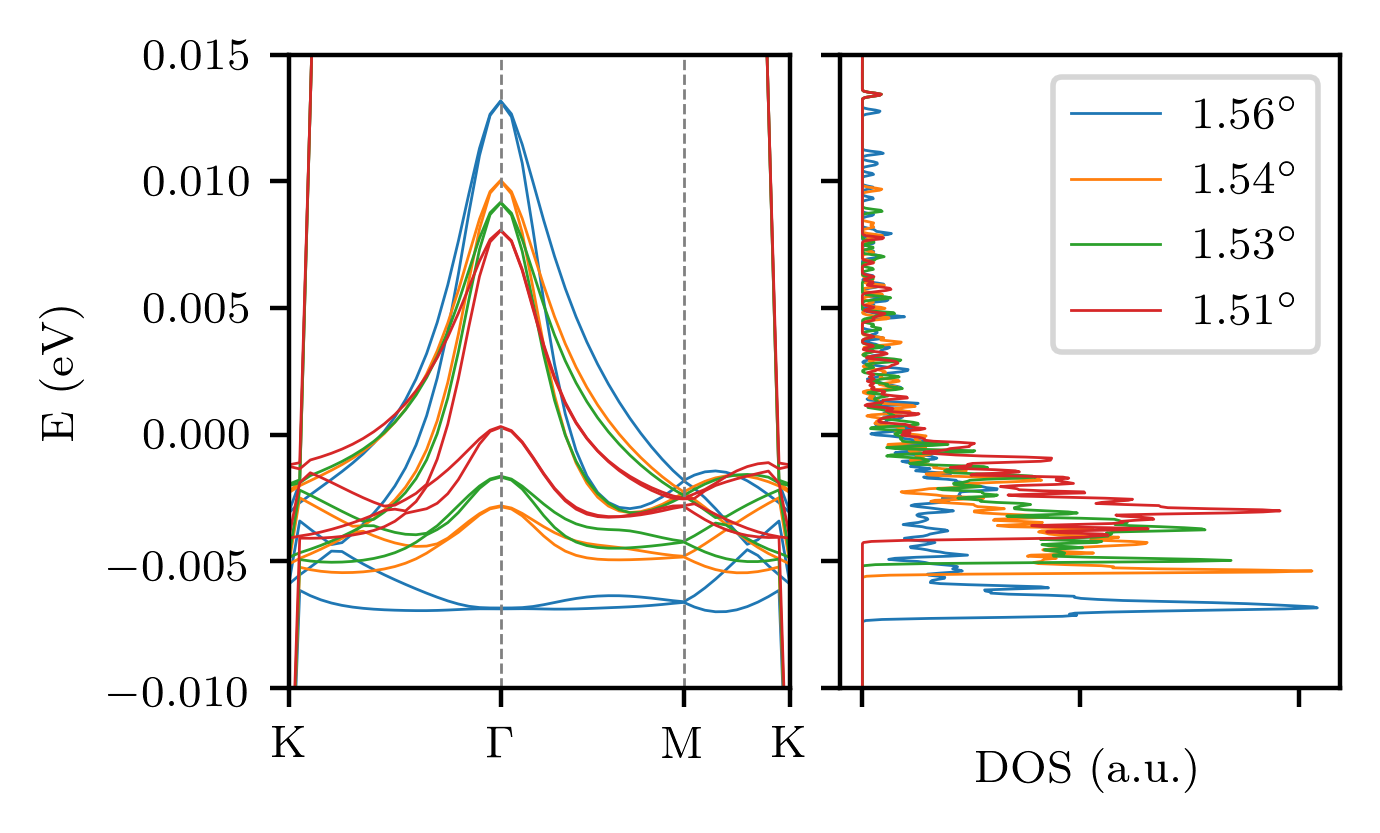}
\caption{(color online) For t3G, we can see the close agreement between the bandstructures and DOS calculations where different maxima are observed depending on which of the flat bands is flatter for a specific angle. The values of $S^\prime$ used in this figure, related to $\theta_\text{eff}$ using Eq.~(\ref{effTheta}), are $0.880$, $0.895$, $0.900$ and $0.914$, respectively. 
For the very small broadening of $\eta=0.1$ meV used here, the $\theta=1.56^\circ$ system corresponding to the largest DOS is different from the magic angle of $1.51^\circ$ based on the minimal band-width of the flat bands at the $\Gamma$-point.}
\label{t3G_DOS_EBS}
\end{center}
\end{figure}
In Fig.~\ref{analyticVSNumeric}, we plotted the numerical predictions of the magic angles based on DOS calculations to avoid the numerical complications of having to use relatively small commensurate cells for electronic band structure calculations. Using Lanczos recursion, we are able to finely scan additional angles around the analytical predictions, here done in steps of approximately $\Delta \theta \sim 0.005^{\circ}$. We then considered two different regimes for the extraction of the flattest band, \textit{i.e.} the largest peak in the DOS, based on the broadening value $\eta$ in Eq.~(\ref{eqDos}). In Fig.~\ref{extractMagicAngles_0_001}, we first consider a relatively large broadening of about $\eta = 3.5$ meV which has the tendency peak at one single value in the DOS. For this regime, we thus report the magic angle as the angle corresponding to this largest DOS value. Then, we also consider a second regime using a numerical broadening of $\eta = 0.175$ meV which allows to resolve all the fine electronic band structure features, similar to what had been reported in Ref.~\cite{1910.12805} for tBG. Under these conditions, we observe a large number of peaks in the DOS, hence we extract the range of magic angles reported in Fig.~\ref{analyticVSNumeric} for smallest broadening to be the one encompassed by the angles corresponding to the outer peaks in the illustrated range with an amplitude above a certain arbitrary threshold. Orange lines indicate these values. We note that unlike for the finite systems, the global peak of the DOS maxima for the bulk seems to be mostly insensitive to the chosen broadening due to the large number of bands contributing concurrently to the broad DOS shown in the side panels of Fig.~\ref{BulkEBS}. 
%
%
%
%
%
%
%
%
%
%
%
%
%
%
%
%
%
To illustrate how these fine features match with specific bands in the electronic band structure and are not rooted in numerical noise, we focus on the t3G system in Fig.~\ref{t3G_DOS_EBS}. Here we have used the commensurate cell at $\theta = 1.538^\circ$ to calculate all the curves, changing the value of $S^\prime$ and finding the corresponding effective angle using Eq.~(\ref{effTheta}). We use exact diagonalization for both the electronic bandstructure and the DOS here with a broadening of $\eta = 0.1$ meV. 
Unlike for the larger broadening value used in the main text, the striking observation here is that using a small broadening value of $\eta = 0.1$ meV the flattest band as predicted by the minimum in bandwidth at $\Gamma$, corresponding to an effective magic angle of $\theta_{\rm eff} = 1.51^\circ$ is different from the flattest as predicted by the maximum in DOS, namely $\theta_{\rm eff} = 1.56^\circ$.  We can relate this angle with the largest DOS peak to the very flat valence band. These observations on the t3G system suggests that for the other t$N$G systems the double or triple peak configuration developing for the smallest broadening probably also corresponds to valence and conduction bands reaching maximum flatness at different angles.
%
We note that this effective angle procedure for small commensurate cells used here in Fig.~\ref{t3G_DOS_EBS} is the same one we used for the band structure calculations in the main text. We summarize in Table~\ref{tab:systemparameters} these different system-specific parameters for the band structures in the main text and provide for reference also the $\theta$-values as predicted by the maximum in DOS observing that the magic angles using either criterion are very similar for the finite systems.

\begin{table}[t]
\begin{center}
\begin{tabular}{lllllll}
\hline
$N$   & $\theta_\text{cell}$    & $\theta$ & $\theta_\text{eff}$  & $S^\prime$  &  $i$ $j$ & $\#$ Atoms \\ \hline
2  & 1.08 & 1.08 & 1.08 & 0.895 & 31 30 & 11164 \\
3  & 1.54 & 1.51 & 1.51 & 0.914  & 21 20  & 8322 \\
4  & 1.79 & 1.73 & 1.72 & 0.932  & 19 18  & 8216 \\
5  & 1.89 & 1.82 & 1.84 & 0.921  & 18 17  & 9190 \\
6  & 2.00 & 1.89 & 1.90 & 0.940  & 17 16  & 9804 \\
10 & 2.00 & 2.01 & 2.03  & 0.881  & 17 16  & 16340 \\
20 & 2.00 & 2.02 & 2.02  & 0.884  & 17 16  & 32680 \\
bulk$_{\theta_1}$ & 2.00 & 1.90 & 1.96 & 0.912  & 17 16  & 3268 \\
bulk$_{\theta_2}$ & 2.00 & NA & 1.6 & 1.119  & 17 16  & 3268 \\ \hline
\end{tabular}
\caption{System-specific parameters where $\theta$ is the experimentally most relevant magic angle predicted by our DOS calculations on very large commensurate cells using Eq.~(\ref{eqDos}) while $\theta_\text{cell}$ are the commensuration cells that were used to perform the band structure calculations using the modified $S^\prime$ values to capture the flattest bands using the criterion that minimizes the bandwidth of the flat bands at the $\Gamma$-point. The resulting angles that are probed based on this criterion are given by $\theta_\text{eff}$ and give almost the same angle values as the $\theta$ ones coming from the maximum of the DOS. The two $S^\prime$ values for the bulk correspond to the two panels from the bulk bandstructure figure, \textit{i.e.} the angles corresponding to the minimization of the first magic angle bandwidth at the $\Gamma_1$ ($\Gamma_2$) points for the first (second) bilayer bands, respectively. $\theta_2$ based on the maximum of DOS is not available (NA) due to being hard to resolve among the many other states present in the bulk.
}
\label{tab:systemparameters}
\end{center}
\end{table}

\section{Impact of relaxation effects on magic angle predictions and higher-order magic angles}
\label{relaxationEffectSect}

\begin{figure}[tbhp]
\begin{center}
\includegraphics[width=0.9\columnwidth]{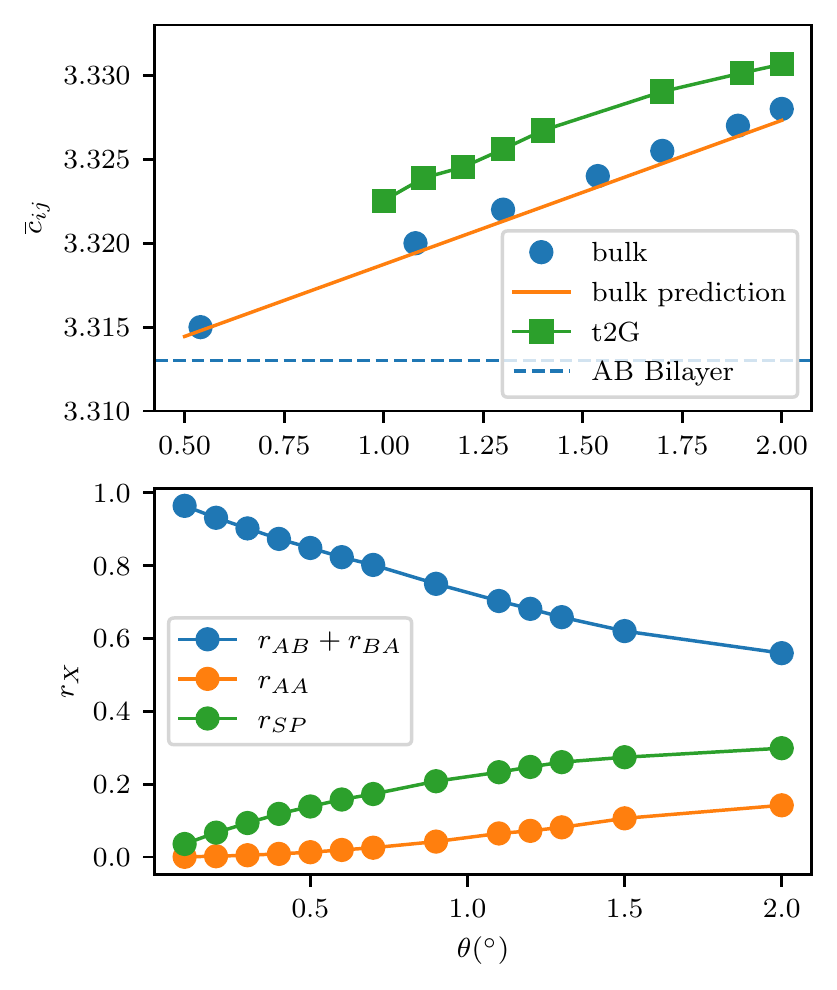}
\caption{(color online) {\em Top panel:} Angle-dependent average interlayer distance $\overline{c}_{ij}$ for the twisted bulk configuration and the free-standing t2G systems. The orange line uses Eq.~(\ref{ratioEq}) to predict the equilibrium bulk interlayer distance using the AA, AB/BA and SP-stacking ratios and their respective equilibrium interlayer distance. We add the equilibrium distance for AB-stacked graphene for reference using the blue-dashed line. {\em Bottom panel:} The numerical data representing the stacking area ratio $r_{X}$ from Eq.~(\ref{ratioEq}) based on free-standing bilayer calculations and used in the top panel for the bulk system based on the stacking-assignment conventions from Ref.~\cite{1910.12805}. These ratios add up to 1 where the moire maps of the smallest angles are almost exclusively formed by the most stable AB and BA stacking regions.}
\label{bulkEqDist}
\end{center}
\end{figure}

In the main text, we elucidated how the in-plane strain plays an important role in explaining the decrease in magic angle observed when compared with the analytical predictions based on the t2G hopping values. Here we provide some additional figures and discussions to go into more detail on these considerations.
In Fig.~\ref{bulkEqDist}, we show how the bulk equilibrium interlayer distance varies with twist angle and how it compares with the average interlayer distance of a t2G system. We notice that this bulk value can almost exactly be predicted using the ratios of AA, AB/BA and SP-stacking present at a specific rotation angle, where we use the conventions outlined in Ref.~\cite{1910.12805} to assign a stacking label to each atom, following
\begin{equation}
    \overline{c}_{ij}^\text{bulk} = r_\text{AA} c_0^\text{AA} + r_\text{AB/BA} c_0^\text{AB/BA} + r_\text{SP} c_0^\text{SP}
    \label{ratioEq}
\end{equation}
where $r_{X}$ corresponds to the ratio of atoms in $X$-stacking with respect to the total number of atoms, while $c_{X}$ is the equilibrium interlayer distance for stacking $X$, equal to $3.313$, $3.402$ and $3.317$\AA\ for AB/BA, AA and SP-stacking, respectively. The free-standing system shows slightly larger values than the bulk since it is not constrained by the periodic boundaries in the vertical direction.
\begin{figure}[tbhp]
\begin{center}
\includegraphics[width=1.0\columnwidth]{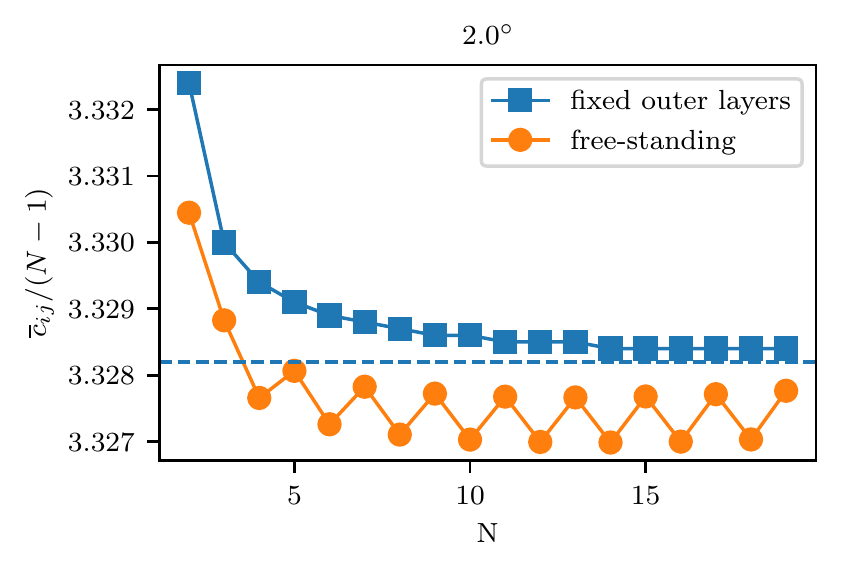}
\caption{(color online) Layer-dependent average interlayer distance normalized by the number of layers for a $2^\circ$ twisted system where we allow for out-of-plane corrugation (orange) and where we preclude this corrugation by fixing the z-coordinate of the outer layers (blue). The free-standing system reaches a minimum around N=14 thus giving a possible reason for the maximum in magic angle at $N=10$ observed in Fig.~\ref{analyticVSNumeric} due to increased tunneling and thus larger magic angle. When fixing the outer-layers, we observe that near-bulk behavior is reached after 5 layers where the interlayer distance departs by $0.001$~\AA\ from the bulk-value for this angle.}
\label{averageDistWithN}
\end{center}
\end{figure}
In Fig.~\ref{averageDistWithN}, we look at the average interlayer distance for both freestanding t$N$G systems as well as their counterparts fixing the outer-layer atoms. We notice that the low-$N$ t$N$G systems show a larger average interlayer distance than the high-$N$ systems which can be rationalized by the fact that these systems contain in relative terms more single-interface layers where the in-plane relaxation is reduced by almost a factor 2 compared to the inner layers that have two interfaces. This reduction of in-plane relaxation reduces the lattice reconstruction effects and in turn increases the average interlayer distance due to a higher proportion of AA-stacking entering Eq.~(\ref{ratioEq}).
%
%
%
%
%
%
%
%
%
%
%

In order to check if the relaxation effects that tend to decrease the analytical predictions also impact the higher-order magic angles, namely the smaller magic angles corresponding to the effective bilayers
with smaller interlayer tunneling, we perform similar DOS calculations on the first magic angle of the second bilayer bands of t4G, t5G and t6G. We obtain values are that are very similar to the analytical predictions reported in Table~\ref{tab:magicangle_analytic}. This stands in contrast with our previous conclusions on the first magic angle of the first bilayer bands stating that the magic angles are reduced due to relaxation effects. By calculating the layer-resolved DOS we observe that the states associated to the flat bands from the second bilayer bands are located preferentially away from the central layers of the system, including a partial occupation on the outer-most layers. These outer-layer carry a smaller amount of in-plane relaxation due to interfacing only with one other layer, hence the expected magic angle reduction is less pronounced.

\begin{figure}[htbp]
\begin{center}
\includegraphics[width=0.8\columnwidth]{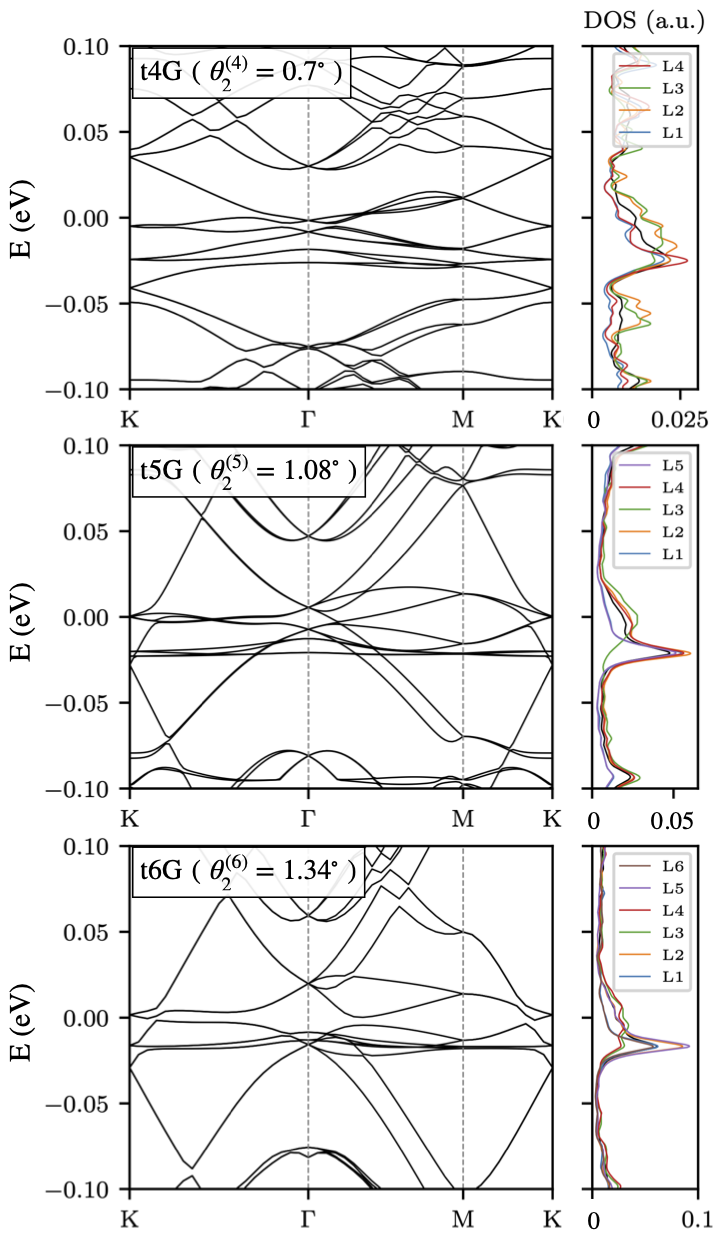}
\caption{(color online) Second magic angle for t4G, t5G and t6G at $\theta = 0.7$, $1.08$ and $1.34^\circ$, respectively, based on the maximum in the DOS criterion. The lowest flat band corresponds to the second bilayer band while the the highest nearly flat band comes from the first bilayer band. The DOS is calculated using a broadening of $\eta=3.5$ meV. Our predictions agree with the analytical predictions in absence of relaxation effects suggesting the latter are less important for these bands than for the first magic angle from the first bilayer bands. This can be rationalized by the fact the states associated with these bands are pushed away from the most inner-layer bands, as illustrated by the layer-resolved DOS, where relaxation effects are weaker. The amplitude of the total density of states in black is divided by the number of layers for comparison purposes.}
\label{2ndMAfor5layer}
\end{center}
\end{figure}

\section{From t$N$G to decoupled t2Gs}
\label{derivationAppendix}
\begin{figure}[tbhp]
\begin{center}
\includegraphics[width=1.0\columnwidth]{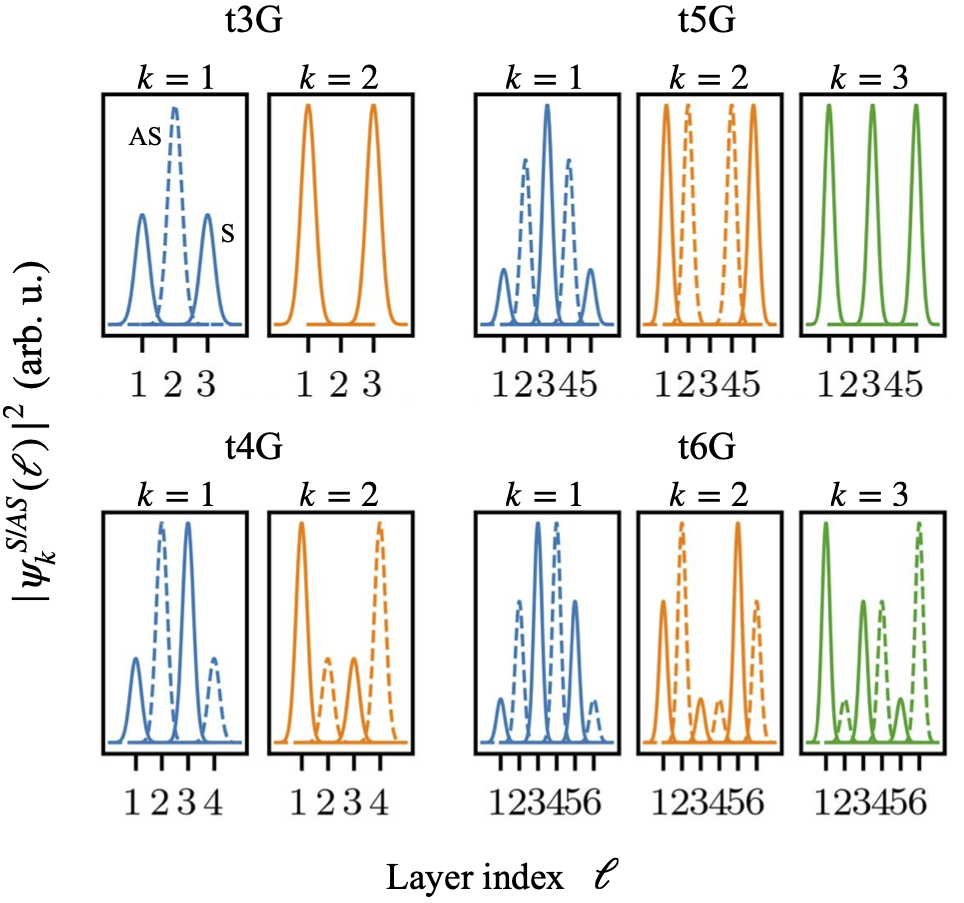}
\caption{(color online) 
Probability distribution of the two wavefunctions with the symmetric(S) and anti-symmetric(AS) combinations of the two states based on the eigenstates with opposite sign eigenvalues  $\lambda_{k} = - \lambda_{N + 1 - k}$ in Eq.~(\ref{1dchainEq}). The states with $k=1$, corresponding to the first magic angle of the first bilayer band in this mapping procedure, are preferentially located on the inner layer(s) while the states next in order with $k=2$, corresponding to the first magic angle of the second bilayer band, get redistributed away from the inner layers where relaxation effects and associated angle reduction effects are weaker when compared to the analytically predicted values based on a constant t2G value.
}
\label{sketch}
\end{center}
\end{figure}
The continuum model Hamiltonian ${\cal H}_{\bm{k}}$ of Eq.~(\ref{Eq:continuumHamil}) can be decoupled the series of decoupled t2G-like Hamiltonian~\cite{khalaf2019} based on singular value decomposition of a layer hopping matrix where odd and even layers were separated. 
Here we show that equivalent results can be obtained by applying in twisted systems the same
basis change and alternating layer/sublattice grouping scheme outlined in Ref.~\cite{koshino2008} for zero twist Bernal stacked graphene multilayers. 

To easily decouple the states, we first rearrange the order of the basis $\{|A_{\ell}\rangle,|B_\ell\rangle\}$ for $\ell=1,2,\cdots,N$ through the transformation matrix $P$ from $\left( \psi_{A_1}, \psi_{B_1}, \psi_{A_2}, \psi_{B_2},\cdots, \psi_{A_N}, \psi_{B_N}\right)^T$ to 
$\left( \psi_{A}^{\rm o}, \psi_{B}^{\rm o}, \psi_{A}^{\rm e}, \psi_{B}^{\rm e}\right)^T$ where
$\psi_{s}^{\rm o} = ( \psi_{s_1}, \psi_{s_3},\cdots,\psi_{s_{2N_o-1}} )^T$ and $\psi_{s}^{\rm e} = ( \psi_{s_2}, \psi_{s_4},\cdots,\psi_{s_{2N_e}} )^T$ 
for the sublattice $s =A, B$ and the number of odd/even layers $N_{o/e}$.
The transformed Hamiltonian from that in Eq.~(\ref{Eq:continuumHamil}) is
\begin{equation}
P^{\dagger}{{\cal{H}}}_{\bm{k}}P \\
= 
\begin{pmatrix}
h_{oo}^{AA}\bm{1}_{o}&h_{oo}^{AB}\bm{1}_o&T_{oe}^{AA}\bm{C_1} &T_{oe}^{AB}\bm{C_1}\\
h_{oo}^{BA}\bm{1}_{o}&h_{oo}^{BB}\bm{1}_o&T_{oe}^{BA}\bm{C_1} &T_{oe}^{BB}\bm{C_1}\\
T_{eo}^{AA}\bm{C_1}^{\dagger}  &T_{eo}^{AB}\bm{C_1}^{\dagger}&h_{ee}^{AA}\bm{1}_e &h_{ee}^{AB}\bm{1}_e\\
T_{eo}^{BA}\bm{C_1}^{\dagger}  &T_{eo}^{BB}\bm{C_1}^{\dagger}&h_{ee}^{BA}\bm{1}_e &h_{ee}^{BB}\bm{1}_e
\end{pmatrix}
\end{equation}
where 
$\bm{1}_{o/e} = \sum_{n=1}^{N_{o/e}} \delta_{n,n}$ are the identity matrices and $\bm{C_1} = \sum_{i=1}^{N_o} [\delta_{n,n} + \delta_{n+1,n}]$ reflects $N_o \times N_e$ interlayer tunneling matrices between the odd and even layers, and $P = \sum_{n=1}^{N_o}\left(\delta_{\,2(2n-1)-1,\,n} + \delta_{\,2(2n-1),\,n+N_o}\right)+ \sum_{n=1}^{N_e}\left(\delta_{\,2(2n),\,n+2N_o} + \delta_{\,2(2n),\,n+2N_o+N_e}\right)$ is the transformation matrix.
For simplicity, we assume zero on-site energy for each layer denoting $h_{oo}^{AA/BB}=h_{ee}^{AA/BB}=0$. 
The $\bm{k}$ vector dependence is only included in $h_{oo}^{AB}\ e^{i\theta/2} = [h_{oo}^{BA}\ e^{-i\theta/2}]^{\dagger} = \hbar\upsilon_F {\pi}_{\bm{k}}^{\dagger}$ and $h_{ee}^{AB}\ e^{-i\theta/2} = [h_{ee}^{BA}\ e^{+i\theta/2}]^{\dagger} = \hbar\upsilon_F {\pi}_{\bm{k}'}^{\dagger}$.
Interlayer tunneling matrices $T_{oe}^{\alpha\beta}(\bm{r})=T_{eo}^{\alpha\beta}(-\bm{r}) = \sum_{j=0,\pm}\ T_j^{\alpha\beta}(\bm{r})$ are the same as in Eq.~(\ref{Eq:tunneling}) where
$$\left(T_j^{AA}({\bf r}),T_j^{AB}({\bf r}),T_j^{BA}({\bf r}),T_j^{BB}({\bf r})\right)^T$$  
$$=e^{-i\,{\bf Q}_j\,\cdot\,{\bf r}} \left(\omega^{\prime},\omega e^{-i\phi_j},\omega e^{+i\phi_j},\omega^{\prime}\right)^T.$$ 

Noting that $\bm{C_1}$ resembles the off-diagonal parts of the 1-dimensional chain system and they can easily be decoupled by even and odd atoms, we adopt the new sublattice $s=A,B$ resolved basis $\left\{\ |\tilde{\psi}_{k}^{\rm odd(A)} \rangle, |\tilde{\psi}_{k}^{\rm odd(B)} \rangle, |\tilde{\psi}_{k}^{\rm even(A)} \rangle, |\tilde{\psi}_{k}^{\rm even(B)} \rangle \right\}$ as in Eq.~(\ref{basis})
where
\begin{equation}
|\tilde{\psi}_{k}^{\rm odd(s)} \rangle=\sum_{\ell=1}^{N} \psi_{k}^{S}(\ell)\ |{s_{\ell}}\rangle,\ 
|\tilde{\psi}_{k}^{\rm even(s)} \rangle=\sum_{\ell=1}^{N} \psi_{k}^{AS}(\ell)\ |{s_{\ell}}\rangle  
\label{basistransform}
\end{equation}
with the symmetric(S)/anti-symmetric(AS) combinations of two wave functions 
for each layer defined as
\begin{equation}
\begin{aligned}
    \psi_{k}^{S}(\ell)&=\frac{\psi_k (\ell) +\psi_{N+1-k} (\ell)}{\sqrt{2}} = 
    &{\sqrt{2}}\ (-1)^{(\ell-1)}\ \psi_k (\ell) \\
    && ({\rm for\ odd}\ \ell) \\
    \psi_{k}^{AS}(\ell)&=\frac{\psi_k (\ell) -\psi_{N+1-k} (\ell)}{\sqrt{2}} = 
     &{\sqrt{2}}\ (-1)^{(\ell-1)}\ \psi_k (\ell) \\
    &&  ({\rm for\ even}\ \ell)
    %
    %
    \end{aligned}
\end{equation}
with $k=1,2,\cdots,N_e$ following Eq.~(\ref{1dchainEq})$-$(\ref{basis}) in the main text.

The final decoupled Hamiltonian is
\begin{equation}
\tilde{{\cal{H}}}_{\bm{k}} = \tilde{P}^{\dagger}P^{\dagger}{{\cal{H}}}_{\bm{k}}P\tilde{P} \\
= 
\begin{pmatrix}
h_{oo}^{AA}\bm{1}_{o}  &h_{oo}^{AB}\bm{1}_o    &T_{oe}^{AA}\bm{\Lambda}&T_{oe}^{AB}\bm{\Lambda}\\
h_{oo}^{BA}\bm{1}_{o}  &h_{oo}^{BB}\bm{1}_o    &T_{oe}^{BA}\bm{\Lambda}&T_{oe}^{BB}\bm{\Lambda}\\
T_{eo}^{AA}\bm{\Lambda}&T_{eo}^{AB}\bm{\Lambda}&h_{ee}^{AA}\bm{1}_e    &h_{ee}^{AB}\bm{1}_e\\
T_{eo}^{BA}\bm{\Lambda}&T_{eo}^{BB}\bm{\Lambda}&h_{ee}^{BA}\bm{1}_e    &h_{ee}^{BB}\bm{1}_e
\end{pmatrix}
\end{equation}
that becomes the series of decoupled t2G-like Hamiltonian
\begin{equation}
\begin{aligned}
{{\cal{H}}}_{\bm{k}}^{\rm (dec)}
=&P \tilde{{\cal{H}}}_{\bm{k}} P^{\dagger} \\ 
=&{\cal{H}}_{\bm{k}\lambda_{1}}^{(t2G)}\oplus {\cal{H}}_{\bm{k}\lambda_{2}}^{(t2G)}\oplus \cdots \oplus {\cal{H}}_{\bm{k}\lambda_{N_e}}^{(t2G)}\\ 
 &\left(\oplus {\cal{H}}_{\bm{k}}^{(1G)}\right)
\end{aligned}
\end{equation}
where 
$\Lambda = \sum_k \ \lambda_{k} \delta_{k,n}$ is the diagonal $N_o \times N_e$ matrices with $k,n=1,2,\cdots, N_e$ with the eigen value $\lambda_{k}$ of 1d chain in Eq.~(\ref{1dchainEq}). 
The transform matrix $\tilde{P} = {\rm diag}\left(\tilde{P}_{\rm o},\tilde{P}_{\rm o},\tilde{P}_{\rm e},\tilde{P}_{\rm e}\right)$ consists of $\left[\tilde{P}_{\rm o/e}\right]_{n,k} = \psi_{k}^{S/AS}(\ell)$ with $k,n=1,2,\cdots,N_{o/e}$, $\ell = 2n$ for even atoms, and $\ell=2n-1$ for odd atoms.

We can write the $k$-th decoupled t2G eigen functions satisfying ${{\cal{H}}}_{\bm{k}}^{\rm (dec)}(\bm{r})\Psi_{\lambda_k}^{\,t2G}(\bm{r}) = \Lambda_{\lambda_k}(\bm{r})\ \Psi_{\lambda_k}^{t2G}(\bm{r})$ in terms of the linear combination of the states for each layer,
\begin{equation}
    \begin{aligned}
    \Psi_{\lambda_k}^{t2G}(\bm{r})
    =&\sum_{\ell=1}^{N}\ \left[\tilde{P}_{\rm o/e}\right]_{n,k}\ \left[\sum_{s=A,B}\ \Psi_{\lambda_k}^{(s_{\ell})}(\bm{r})\right]\\
    =& \sum_{\ell=1}^N\ 
\left[ {\sqrt{2}}\ (-1)^{(\ell-1)}\ 
    \sqrt{\frac{2}{N+1}}\sin \left( \kappa_k\ \ell \right) \right] 
\ \Psi_{\lambda_k}^{(\ell)}({\bm{r}}),
    \end{aligned}
\end{equation}
for $k=1,2,\cdots,N_e$.

In order to understand the spatial distribution of states as shown the far-most right panels in Fig.~\ref{relaxationEffect}, we use this decoupling procedure to confirm in Fig.~\ref{sketch} that the highest eigenenergies with value $k=1$ in Eq.~(\ref{1dchainEq}) shows the same distribution of states as the first magic angle shown from Fig.~\ref{relaxationEffect} and the states with index $k=2$ match the distribution of states from the layer-resolved DOS at the energy from the second magic angle in Fig.~\ref{2ndMAfor5layer}.
%

\begin{figure}[tbhp]
\begin{center}
\includegraphics[width=0.8\columnwidth]{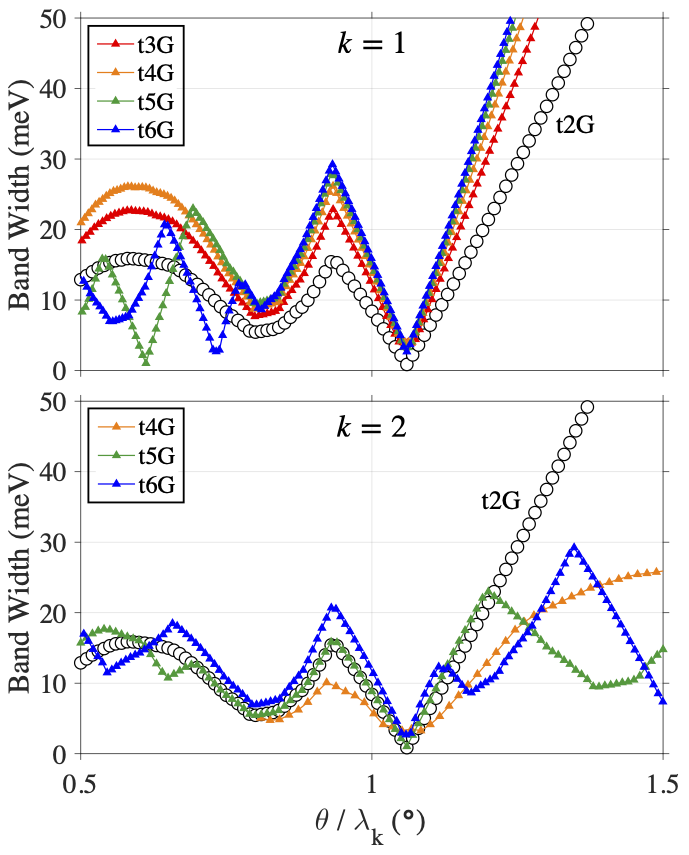}
\caption{(color online) Band width profile of t2G$\sim$t6G   calculated by a continuum model as a function of twist angle $\theta$. 
The magic angle of t$N$G at which the band widths are minimized coincides with the first magic angle of t2G with the factor $\lambda_k = 2\cos\left(\frac{\pi k}{N+1}\right)$ in Eq.~(\ref{1dchainEq}).}
\label{continuum_bandwidth}
\end{center}
\end{figure}

\label{continuumAppendix}
We further provide numerical confirmation of the decoupling procedure by illustrating how the scaling factor $\lambda_k$ from Eq.~\ref{1dchainEq} allows the first magic angle of t$N$G to coincide with the first magic angle of t2G. This is illustrated in Fig.~\ref{continuum_bandwidth} for $\lambda_1$ and $\lambda_2$. This figure illustrates that the analytical predictions are quite accurate using the continuum model but based on the observations drawn from Fig.~\ref{analyticVSNumeric} we expect a departure from this ideal trend when realistic strain profiles are included.
Table~\ref{tab:magicangle_cont} summarizes the first magic angles for t2G all the way to the bulk obtained numerically using the continuum model, while Table~\ref{tab:magicangle_analytic} provides the $\theta_i$ values corresponding to the so called chiral model's $\alpha_i$ values from Ref.~\cite{khalaf2019} taking as t2G reference the value of $\theta^{(2)}_1=1.08^{\circ}$.
%

\begin{table}[tbhp]
\begin{center}
\begin{tabular}{ll|ll}
\hline
$N$ & $\theta^{(N)}_1$ & N & $\theta^{(N)}_1$\\ \hline
2 & 1.06 & 8 & 2.00 \\\cline{1-2} 
3 & 1.50 &   & 1.62 \\\cline{1-2} 
4 & 1.71&    & 1.06 \\\cline{3-4} 
  & 0.67 & 10& 2.03 \\\cline{1-2} 
5 & 1.84&    & 1.78 \\
  & 1.06 &   & 1.39 \\\cline{1-2} 
6 & 1.91&    & 0.88 \\ \cline{3-4}
  & 1.32 & 20& 2.10 \\ \cline{3-4}
  & 0.47 & bulk& 2.12\\\hline
\end{tabular}
\caption{The magic angles for t$N$G numerically obtained by a continuum model with constant $\omega=0.12~\rm{eV}$, and $\omega^{\prime}=0.0939~\rm{eV}$.}
\label{tab:magicangle_cont}
\end{center}
\end{table}

\begin{table}[tbhp]
\begin{center}
\begin{tabular}{lllllll}
\hline
N & $\theta^{(N)}_1$  & $\theta^{(N)}_2$  & $\theta^{(N)}_3$ & $\theta^{(N)}_4$ & $\theta^{(N)}_5$ & $\theta^{(N)}_6$ \\
\hline
2 & 1.080 & 0.285 & 0.169 & 0.120 & 0.093 & 0.076 \\ \hline
3 & 1.528 & 0.246 & 0.239 & 0.170 & 0.132 & 0.108 \\ \hline
4 & 1.748 & 0.461 & 0.273 & 0.194 & 0.151 & 0.123 \\
  & 0.668 & 0.176 & 0.104 & 0.074 & 0.058 & 0.047 \\ \hline
5 & 1.872 & 0.494 & 0.292 & 0.208 & 0.161 & 0.132 \\
  & 1.080 & 0.285 & 0.169 & 0.120 & 0.093 & 0.076\\ \hline
6 & 1.947 & 0.513 & 0.304 & 0.216 & 0.168 & 0.137 \\
  & 1.346 & 0.355 & 0.210 & 0.150 & 0.116 & 0.095 \\
  & 0.480 & 0.127 & 0.075 & 0.053 & 0.041 & 0.034\\
 \hline
\end{tabular}
\caption{
Magic angle families $\theta^{(N)}_1\sim\theta^{(N)}_6$ in the chiral limit ($\omega^{\prime}=0$)
can be converted from the $\alpha_1\sim\alpha_6$ values from Ref.~\cite{khalaf2019} using Eq.~(\ref{hierarchy}) 
with $\upsilon_F = (\sqrt{3}a/2\hbar) |t_0| \approx 1\times 10^6~\rm m/s$ with a nearest hopping parameter $t_0=-3.1~\rm eV$, the intersublattice interlayer coupling $\omega_{AB^{\prime}} = 0.1242~\rm eV$, and $k_D = 4\pi/3a$ with a lattice constant $a=2.46$~\AA. 
}
\label{tab:magicangle_analytic}
\end{center}
\end{table}

\begin{figure*}[htbp]
\begin{center}
\includegraphics[width=1.0\textwidth]{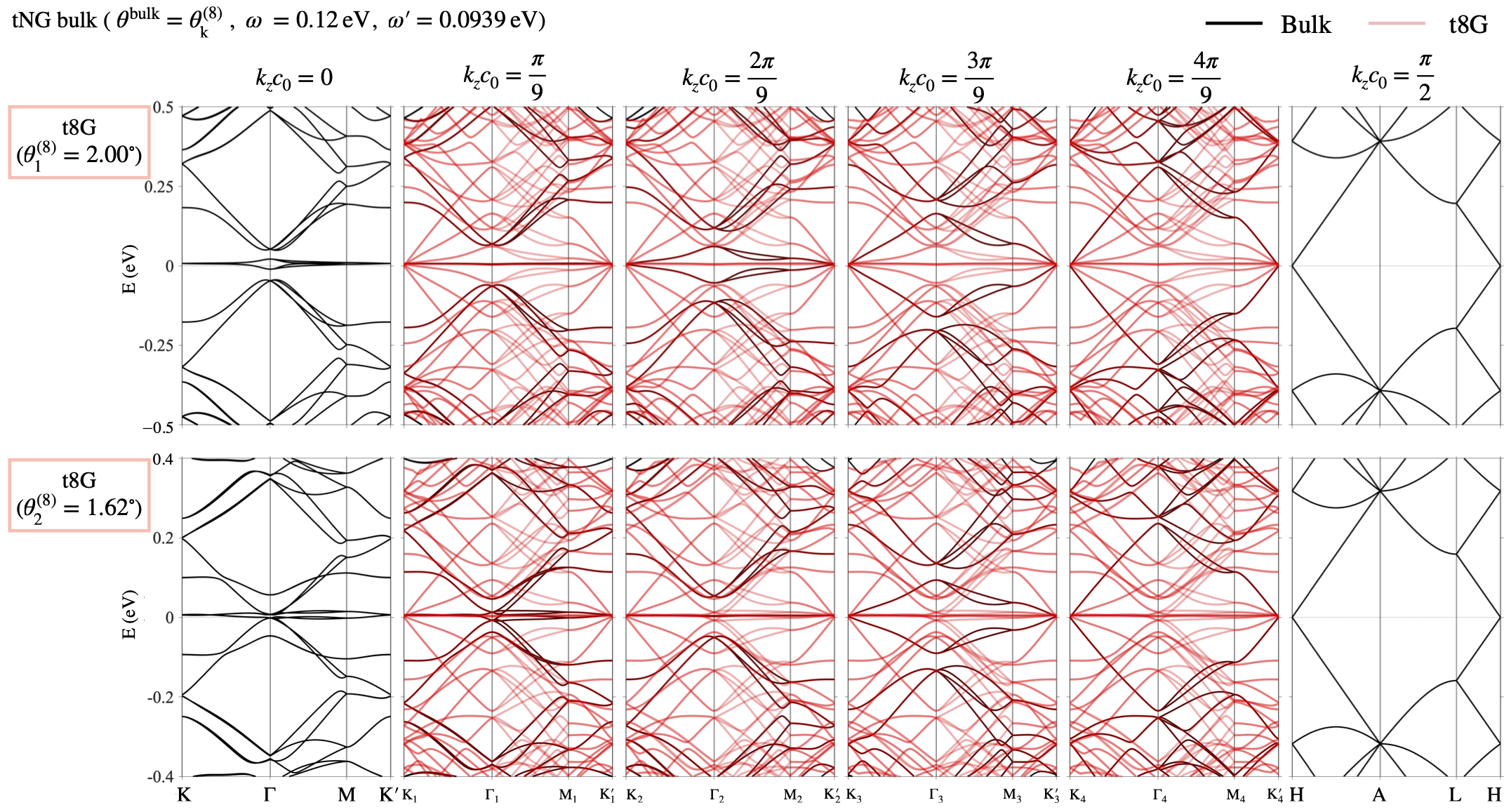}
\caption{(color online) Bulk (black) and t8G (red) electronic band structures using the continuum model where we tune $\theta_1^{(8)}=2.00^{\circ}$, and $\theta_2^{(8)}=1.62^{\circ}$ to the first and second magic angle of t8G, and $\theta^{\rm bulk}=\theta_k^{(8)}$ for the bulk. Similar to Fig.~\ref{BulkEBS}, this allows us to probe both the first and second magic angle in the $K, \Gamma, M, K^{\prime}$ plane where in both cases the bulk to finite mapping matches reasonably well. The tunneling amplitudes used for $\omega$ and $\omega^{\prime}$ are the same in Fig.~\ref{relaxationEffect}, and the height of a unit cell of the bulk in $z$-direction is $2c_0$.
} 
\label{continuumBulk}
\end{center}
\end{figure*}

%
We finally illustrate in Fig.~\ref{continuumBulk} a similar finite-to-bulk bandstructure calculation as in Fig.~\ref{BulkEBS}, but using the continuum model. We observe perfect overlap at the selected cuts for the bulk with the corresponding t8G bilayer bands as the relaxation profiles are exactly the same by design when using the continuum model.

\end{document}